\documentclass[lettersize,journal]{IEEEtran}
\usepackage{amsmath,amsfonts}

\usepackage{algorithmic}
\usepackage{hyperref}
\usepackage{algorithm}
\usepackage{array}
\usepackage{subfigure}
\usepackage{textcomp}
\usepackage{stfloats}
\usepackage{url}
\usepackage{verbatim}
\usepackage{graphicx}
\usepackage{cite}
\usepackage{braket}
\usepackage{multirow}
\usepackage{multicol}
\usepackage[square, comma, sort&compress, numbers]{natbib}
\usepackage{amssymb}
\usepackage{hhline}

\usepackage{makecell}
\usepackage{gensymb}
\usepackage[normalem]{ulem}
\useunder{\uline}{\ul}{}
\begin{document}

\title{Adaptive Quantum Scaling Model for Histogram Distribution-based Quantum Watermarking}

\author{Zheng Xing,
Chan-Tong Lam,~\IEEEmembership{Senior Member,~IEEE,}
Xiaochen Yuan*,~\IEEEmembership{Senior Member,~IEEE,} 
Sio-Kei Im,
Penousal Machado

\thanks{This work was supported in part by the Macao Polytechnic University under grant RP/FCA-04/2024, and the Science and Technology Development Fund of Macau SAR under grant 0045/2022/A. (Corresponding author: X. Yuan)}
\thanks{Z. Xing, C.T. Lam, and X. Yuan are with the Faculty of Applied Sciences, Macao Polytechnic University, Macau S.A.R 999078, China (e-mail: zheng.xing@mpu.edu.mo; ctlam@mpu.edu.mo; xcyuan@mpu.edu.mo)}
\thanks{S.K. Im is with the Macao Polytechnic University, Macau S.A.R 999078, China (email: marcusim@mpu.edu.mo)}
\thanks{P. Machado is with the Department of Informatics Engineering, University of Coimbra, Coimbra 3004530, Portugal (e-mail: machado@dei.uc.pt)}}

\markboth{Journal of \LaTeX\ Class Files,~Vol.~14, No.~8, August~2021}%
{Shell \MakeLowercase{\textit{et al.}}: Adaptive Quantum Scaling Model for Histogram Distribution-based Quantum Watermarking}

\IEEEpubid{0000--0000/00\$00.00~\copyright~2021 IEEE}

\maketitle

\begin{abstract}
The development of quantum image representation and quantum measurement techniques has made quantum image processing research a hot topic. In this paper, a novel Adaptive Quantum Scaling Model (AQSM) is first proposed for scrambling watermark images. Then, on the basis of the proposed AQSM, a novel quantum watermarking scheme is presented. Unlike existing quantum watermarking schemes with fixed embedding scales, the proposed method can flexibly embed watermarks of different sizes. In order to improve the robustness of the watermarking algorithm, a novel Histogram Distribution-based Watermarking Mechanism (HDWM) is proposed, which utilizes the histogram distribution property of the watermark image to determine the embedding strategy. In order to improve the accuracy of extracted watermark information, a quantum refining method is suggested, which can realize a certain error correction. The required key quantum circuits are designed. Finally, the effectiveness and robustness of the proposed quantum watermarking method are evaluated by simulation experiments on three image size scales. The results demonstrate the invisibility and good robustness of the watermarking algorithm.

\end{abstract}


\section{Introduction}
\IEEEPARstart{Q}{uantum} image technology is one of the high-profile fields in recent years, which applies the potential quantum computing power \citep{tan2020optimality,el2023towards} to the image processing field. With the development of quantum information technology, quantum image processing research \citep{liu2022quantum} has become a hot topic. In the classical field, the watermarking technology proves the copyright \citep{hua2022matrix} and authenticity of digital media by embedding specific identification information in the digital media, which can be detected and extracted. Resembling classical digital watermarking techniques \citep{yuan2018local,yuan2021gauss}, quantum watermarking schemes aim to preserve the copyrights of quantum images by embedding invisible watermark into carrier images for verifying its ownership \citep{xing2023novel}. Since the mechanisms of quantum computers \citep{munoz2018quantum,verma2025quantum,tennie2025quantum} are different from electronic computers, to utilize quantum computing, quantum image processing methods need to be developed. A prerequisite for quantum image techniques is the representation and storage of images on a quantum computer, and research on quantum image representation has provided several options \citep{venegas2010processing,le2011flexible,zhang2013neqr,xing2024ngqr}.
\par Quantum watermarking methods are classified according to their embedding domains, i.e. the spatial domain based \citep{luo2018enhanced,hu2019lsbs} and the frequency domain based \citep{heidari2017novel}. The former embeds the watermark in the spatial domain, while the latter converts the carrier image into the frequency domain and then embeds the watermark into frequency domain. In 2013 and 2014, based on the Flexible Representation of Quantum Image (FRQI), quantum watermarking schemes was proposed \citep{zhang2013quantum,song2013dynamic,song2014dynamic}. However, the pixel color information of FRQI images is encoded by the amplitude of a single qubit. Therefore, the methods in \citep{zhang2013quantum,song2013dynamic,song2014dynamic} embedded the pixel information of the quantum watermark image directly to the carrier image information, which violates the principles of quantum mechanics.  It should be noted that the quantum watermarking algorithm based on the frequency domain is not accurate, which was confirmed in \citep{yang2013analysis,yang2016letter}. In 2015, Yan et al \citep{yan2015duple} proposed a dual watermarking strategy for multichannel quantum images. This strategy embedded the watermark image information in both spatial and frequency domains. Two keys were generated using quantum measurements and owner assignment respectively during preprocessing of the watermark. They were used to scramble the watermark during the embedding process and to recover the watermark when extracted. In 2016, Miyake and Nakamae \citep{miyake2016quantum} proposed a new quantum grayscale image watermarking scheme using a simple small-scale quantum circuit by employing the Novel Enhanced Quantum Image Representation (NEQR).\IEEEpubidadjcol First, the grayscale watermark was extended to a binary grayscale image. Then, the binary image was scrambled by SWAP gate \citep{ono2017implementation}. Finally, the embedding was implemented by XOR operation. However, only the robustness under noise addition was analyzed and more attacks were not considered. In 2017, for quantum color images, Li et al. \citep{li2017improved} proposed an improved scheme using small-scale quantum circuits and pixel-value scrambling. First, the color of the pixels in the watermarked image is scrambled using a controlled rotation gate, and then the watermark is expanded. Finally, embedding is achieved by CNOT gate. In the same year, Naseri et al. \citep{naseri2017new} proposed a new quantum image watermarking strategy that employs the most significant bit (MSB) in addition to the least significant bit (LSB). This scheme improves the noise immunity but only gives limited data under noise addition without considering other attacks.
In 2018, Luo et al \citep{luo2018enhanced} proposed an enhanced quantum watermarking scheme. Since human vision is not sensitive to the edge region of the image, the watermark image is embedded into the edge region of the carrier image by LSB substitution in their scheme. it shows high visual effect. However, the payload of this scheme is very low, only 1/16, and there is no scrambling of the watermark image without improved robustness.
In 2019, Luo et al \citep{luo2019adaptive} proposed an adaptive LSB quantum watermarking method using three-way pixel value difference. They segmented the quantum carrier image into $2\times2$ blocks. The blocks were labeled as smooth regions or edge regions by calculating the three-way pixel value difference. The processed quantum watermark information was then embedded into the quantum carrier image by $k$-LSB substitution method. More secret information was embedded in the edge region and less in the smooth region to maintain the visual quality. In 2021, in order to improve the watermark embedding capability and security, Zeng et al. \citep{zeng2021quantum} proposed an improved quantum watermarking algorithm based on maximum pixel difference and tent map. The method is similar to \citep{luo2019adaptive}. The difference is that their scrambling method employs a tent map, which utilizes the maximum difference of four pixel values to determine whether it is a smooth block or an edge block. This results in less visual interference with the watermarked image, but in terms of robustness, only experimental results against noise are provided, and no additional attacks are considered to evaluate robustness. In conclusion, current quantum watermarking schemes tend to fix the size ratio between the embedded watermark image and the carrier image, which is obviously inflexible, and even so, the robustness still needs to be improved. In this paper, to improve the adaptivity and robustness of watermarking algorithms, we first propose a novel Adaptive Quantum Scaling Model (AQSM) for quantum watermarking schemes, which is used to transform a quantum grayscale image into binary images. Subsequently, a novel histogram distribution-based watermarking
mechanism (HDWM) is proposed. Finally, a quantum watermarking scheme utilizing AQSM and HDWM is demonstrated, and simulation experiments are conducted to verify the effectiveness and robustness of the proposed method. Three watermarking algorithms with different embedding ratios (the size ratio of the watermark image to the carrier image) are simulated using database USC-SIPI.

\section{Preliminaries}\label{part2}
\textbf{Quantum Gate}: 
Quantum unitary operators, which are created with quantum gates, are used to implement quantum computation. A unitary matrix can be used to describe quantum operations on a single qubit or multiple qubits \citep{fouskidis2020reconfigurable}. Pauli operators (referred to as $I$, $X$, $Y$, $Z$) and Hadamard transforms are two of the most popular single quantum bit gates (denoted as H). The most frequent X gate is known as the NOT gate, and it is represented by the symbol `$\bigoplus$' in quantum circuits.
The Controlled-NOT (CNOT) gate and the Toffoli gate (also known CCNOT) are important two-qubit and three-qubit gates, respectively. Similarly, SWAP and Fredkin gates (also known as CSWAP) \citep{chattopadhyay2011all} are commonly used two and three qubits gates. 
It is worth emphasizing that single-qubit and two-qubit gates are considered universal, which means they can be used to perform any unitary operation on arbitrary quantities of n-qubits. As a result, a more complex multi-qubit gate can be decomposed into basic quantum logic gates \citep{barenco1995elementary}.

\textbf{NEQR}: 
The NEQR method \citep{zhang2013neqr} uses the base state of a sequence of qubits to store quantum image information, i.e., pixel's grayscale value and corresponding coordinate, into normalized entangled qubits sequence. Representation of NEQR is shown as follows for a grayscale image of size $2^n\times 2^n$ in the range $\left [ 0,2^{q}-1  \right ] $, where $ C_{YX}^{i}\in \left \{ 0,1 \right \} $.
\begin{equation}
    \left | I  \right \rangle=\frac{1}{2^n}\sum_{Y=0}^{2^n-1}\sum_{X=0}^{2^n-1} \otimes _{q-1}^{i=0}\left | C_{YX}^{i} \right \rangle\left | YX \right \rangle      
\end{equation}
The pixel grayscale value $\left | C_{YX}^{i} \right \rangle$ is encoded by a sequence of $q$ qubits, e.g., eight qubits are used for encoding the pixel values of 8-bit grayscale images. The coordinate information is indicated by two sequences of $n$ qubit, denoted as $    \left | Y  \right \rangle=\left | y_{n-1}y_{n-2}...y_{0}  \right \rangle, \left | X \right \rangle=\left | x_{n-1}x_{n-2}...x_{0}  \right \rangle$, where $\ket{y_{i}}, \ket{x_{i}}\in \left \{\ket{0},\ket{1} \right \}$, and $Y,X\in \left [ 0,2^{n}-1  \right ]$. Consequently, for a $2^n\times2^n$ quantum grayscale image, $2n+q$ qubits are required.

\section{The proposed Adaptive Quantum Scaling Model (AQSM)}\label{part3}
\begin{figure*}[h]
	\centering
		\includegraphics[scale=.6]{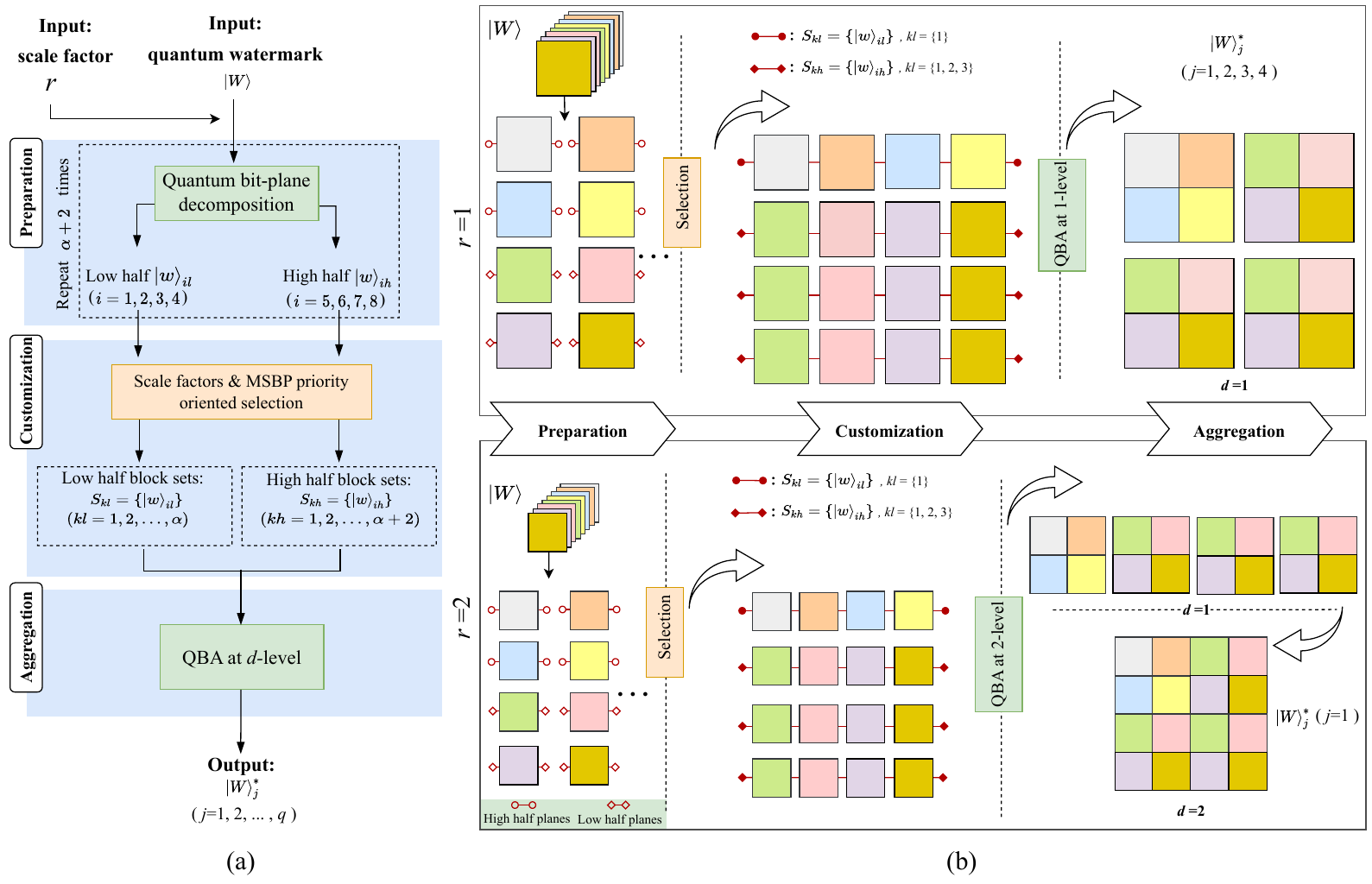}
	\caption{ Overview of the Adaptive Quantum Scaling Model (AQSM). (a) Architecture of AQSM, (b) the demonstrations of $r=1$ and $r=2$.}
	\label{FIG:ATSA_process}
\end{figure*}

In order to expand the grayscale image size and make its pixel values information uniformly distributed, we propose the AQSM, which can transform a quantum grayscale image of size $2^{m}\times2^{m}$ into one or more expanded binary images with no loss of pixel values information. Figure \ref{FIG:ATSA_process} shows the flowchart and demonstrations of the AQSM, where the left side is the flowchart with an input of a quantum watermark image $\ket{W}$ and a scale factor $r$ ($r \in Z^{+}$), and $r$ determines the size of the binary image resulting from the watermark $\ket{W}$ processed by AQSM. The right side shows two examples with scale factors of 1 and 2.
\par
\begin{figure}[h]
	\centering
		\includegraphics[scale=.43]{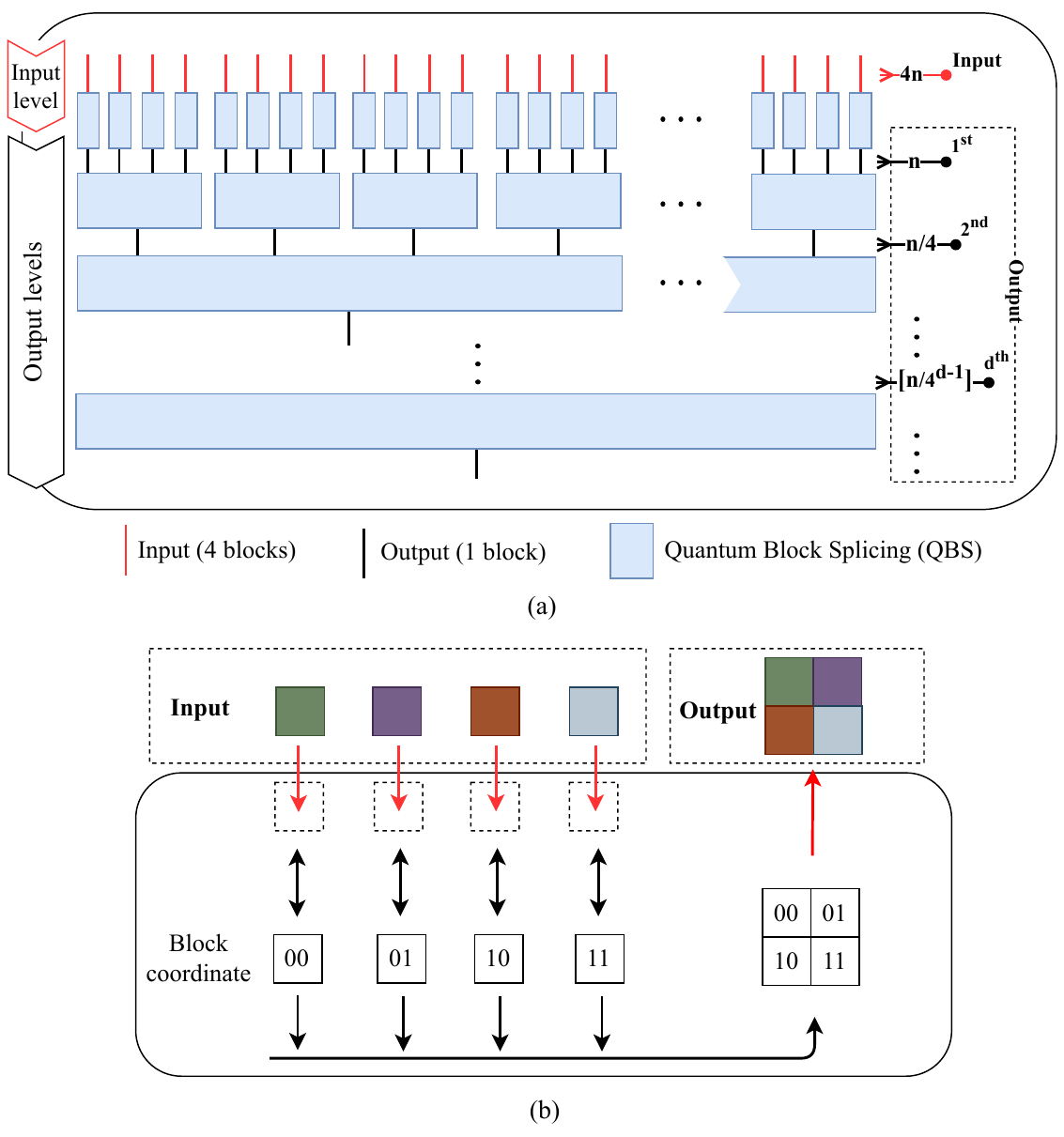}
	\caption{Demonstration of proposed Quantum Block Aggregation (QBA). (a) Overview of QBA, (b) Quantum Block Splicing (QBS) transformation.}
	\label{FIG:QBA_model_demo}
\end{figure}

\begin{figure}[h]
\centering	
\subfigure[]{
		\includegraphics[scale=.4]{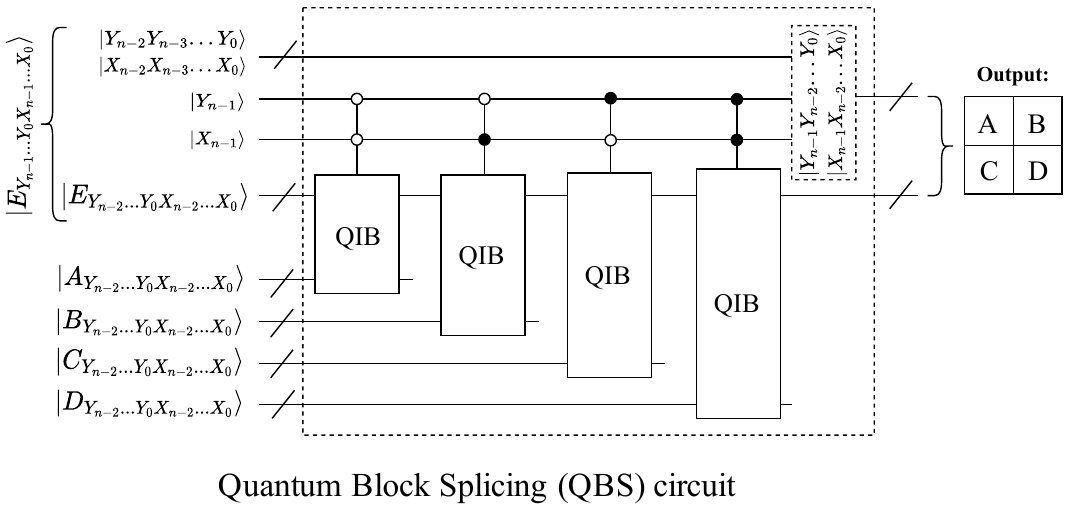}}
  \subfigure[]{
		\includegraphics[scale=.4]{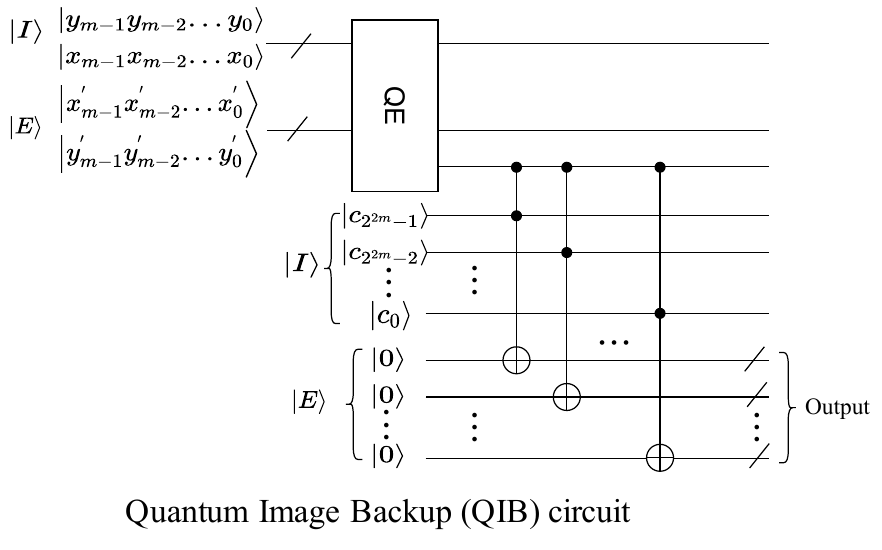}}
	\caption{Quantum circuits of Quantum Block Splicing (QBS). (a) QBS circuit, (b) QIB circuit.}
	\label{FIG:QBS_CIRCUIT}
\end{figure}

 Specifically, the process of AQSM includes three parts: preparation, customization, and aggregation. First, the preparation is achieved by bit-plane decomposition. Since the unknown quantum state cannot be copied, the bit decomposition needs to be repeated $\alpha+2$ times to satisfy the bit-plane needs. The calculation of $\alpha$ is $2^{2\beta-3}-1$, where $\beta$ is defined as follows.
\begin{equation}
    \beta=\left\{
\begin{aligned}
&2, &  & r=1 \\
&r, &  & r>1 
\end{aligned}
\right.
\end{equation}
Note that the eight bit planes obtained from each decomposition are divided into two parts, the low half $\ket{w}_{il}(i=1,2,3,4)$ and the high half $\ket{w}_{ih} (i=5,6,7,8)$. Then, customization is achieved by scale factor \& Most Significant Bit Plane (MSBP) priority oriented selection, which yields the high half block sets denoted as $S_{kl}={\ket{w}_{il}}(kl=1,2,...,\alpha)$ and the low half block sets denoted as $S_{kh}={\ket{w}_{ih}}(kl=1,2,...,\alpha+2)$. And the maximum values of $kl$ and $kh$ depend on $\alpha$. Finally, aggregation is implemented using the QBA module, a demonstration of which is given in Figure \ref{FIG:QBA_model_demo}. 
The overview of QBA is shown in Figure \ref{FIG:QBA_model_demo}(a), including the input level and output level. The red vertical line indicates the input of four blocks, the blue rectangle indicates the Quantum Block Splicing (QBS), and the black vertical line indicates an output. We specify the number of input blocks of the QBA module as $4n$ and the output level is noted as $d(d\le \beta)$, then the output at level $d$ results in a binary image of size $2^{m+d}\times2^{m+d}$ with the number $n/4^{d-1}$. Figure \ref{FIG:QBA_model_demo}(b) shows the interpretation of QBS for four input blocks stitched into an expanded new image according to the corresponding four coordinates 00, 01, 10, and 11. Figure \ref{FIG:QBS_CIRCUIT} shows the quantum circuits for QBS, where the Quantum Image Backup (QIB) is a transfer backup of a quantum image, and its circuit is shown in Figure \ref{FIG:QBS_CIRCUIT}(b). In addition, the Quantum Equal (QE) \citep{zhou2017quantum} uses a qubit to indicates whether two numbers are equal or not. The output indicates the result of the comparison: If the output is 1, it means the two numbers are equal, otherwise they are not equal.

For the QBA module in the AQSM, we set $n=2\alpha+2$, and after the processing of the QBA module at level $d$, the output is marked as $\ket{W}^{\ast}_{j} (j=1,2,...,q)$, where $q$ are defined below.
\begin{equation}
    q=\left\{
\begin{aligned}
&4, &  & r=1 \\
&4^{r-d}, &  & r>1 
\end{aligned}
\right.
\end{equation}
where $r$ is scale factor, $d$ is the level of QBA module. To understand the AQSM intuitively, the processing of r=1 and 2 is shown in the black box in Figure \ref{FIG:ATSA_process}. It can be seen that when $r=1$ the final result is four expanded binary images, and when $r=2$ the output result has two choices, i.e., four expanded binary images at $d=1$ and one expanded binary image at $d=2$, respectively.
\section{Proposed histogram distribution-based quantum watermarking}\label{part4}

The proposed quantum watermarking scheme aims to embed a watermark image of size $2^{m}\times2^{m}$ into a carrier image of size $2^{n}\times2^{n}$, both of which are quantum grayscale images represented by NEQR. First, the required parameters are calculated based on the watermark image and the carrier image, and then the proposed AQSM is applied to generate the reconstructed binary watermarks. Finally, the proposed novel embedding mechanism is utilized to embed the watermark information uniformly into the quantum carrier image resulting in a quantum watermarked image. The extraction process of the quantum watermark is similar to the reverse process of embedding, where the difference is that the proposed quantum refinement method is incorporated to improve the correctness of the extracted watermark information. Finally, the quantum watermark is recovered by bit-plane reconstruction, when the quantum watermarked images need to be visible, only quantum measurements are required. To describe the proposed quantum watermarking scheme more clearly, we first introduce the proposed HDWM as it is used for embedding and extraction. Then the complete process of quantum watermarking is described.
\subsection{Proposed HDWM}

An image histogram \citep{lu2019secure} is a histogram that represents the distribution of brightness in a digital image by plotting the number of pixels for each pixel value in the image. Usually, the left side of the horizontal coordinate of the histogram is the darker area, while the right side is the brighter area. Based on this property, we define $h (g)$ as the number of pixels with a values of $g$ in the histogram distribution of a watermark image. And the cumulative histogram $H(g)$ is refers to the total number of pixels with values not greater than $g$, and its formula is as follows.
\begin{equation}
    H(g)= {\textstyle \sum_{x=0}^{g}}h(x)
\end{equation}
In this paper, we divide pixel values into dark interval ranging from 0 to 127 and bright interval ranging from 128 to 255. The cumulative distribution probability of the dark interval is denoted as $T_{dark}$, while that of bright interval is denoted as $T_{bright}$, and they are calculated as follows.
\begin{equation}
      T_{dark}=\frac{H(127)}{2^m\times 2^n}   
\end{equation}

\begin{equation}
    T_{bright}=1-\frac{H(127)}{2^m\times 2^n}
\end{equation}

Next, we use $T_{dark}$ and $T_{bright}$ of a watermark image to determine the embedding method. First, the division threshold is denoted by $\lambda$, whose values range is from 0 to 1. Then, the absolute values of the difference between $T_{dark}$ and $T_{bright}$ is compared with $\lambda$ to get the embedding parameters $\tau_1$ and $\tau_2$, whose rules are shown in Eq.(5). If the difference is less than $\lambda$, set $\tau_1$ to 0, and vice versa, set 1. If $\tau_2$ is 1, then set $\tau_2$ according to the values of $T_{dark}$ and $T_{bright}$. When $T_{dark}$ is greater than $\frac{1+\lambda}{2}$, set $\tau_2$ to 0, and when $T_{bright}$ is greater than $\frac{1+\lambda}{2}$, set $\tau_2$ to 1.
\begin{equation}
          \tau_{1}=\left\{
\begin{aligned}
    &0, &  & \left | T_{bright}-T_{dark} \right |< \lambda \\
&1, &  & \left | T_{bright}-T_{dark} \right |\ge \lambda 
\end{aligned}
\right.
\end{equation}

\begin{equation}
          \tau_{2}=\left\{
\begin{aligned}
  &0, &  & T_{dark} \ge \frac{1+\lambda}{2} \\
&1, &  & T_{bright} \ge \frac{1+\lambda}{2} 
\end{aligned}
\right.
\end{equation}

After obtaining the embedding parameters according to the histogram distribution, we implement the embedding by a XOR operation, so we define a XOR index $\eta$ to indicate the implementation of the embedding based on XOR result of multi MSB $V_{\eta}$, whose expression goes as follows.
\begin{equation}
    \eta=\left\{
\begin{aligned}
&0, &  & r=1 \\
&r\bmod 2, &  & r>1 
\end{aligned}
\right.
\end{equation}
When $\eta$ is 0, the embedding procedure is indicated using the XOR result of 3 MSBs, and when $\eta$ is 1, the embedding procedure is indicated by the XOR result of 4 MSBs.

\subsection{Quantum watermark embedding using HDWM}
\begin{figure}[h]
	\centering		\includegraphics[scale=.5]{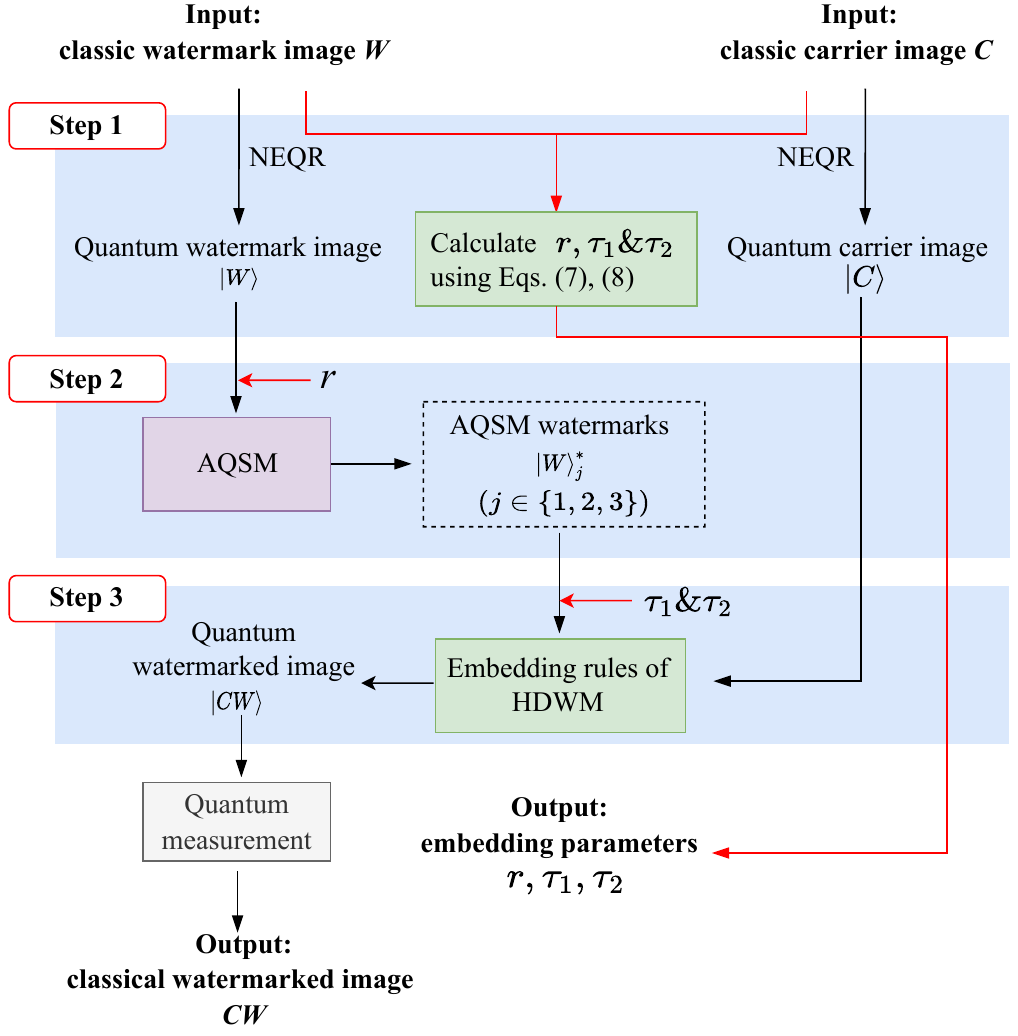}
	\caption{Flowchart of quantum watermark embedding.}
	\label{FIG:embed}
\end{figure}

\begin{figure}[h]
	\centering
		\includegraphics[scale=.43]{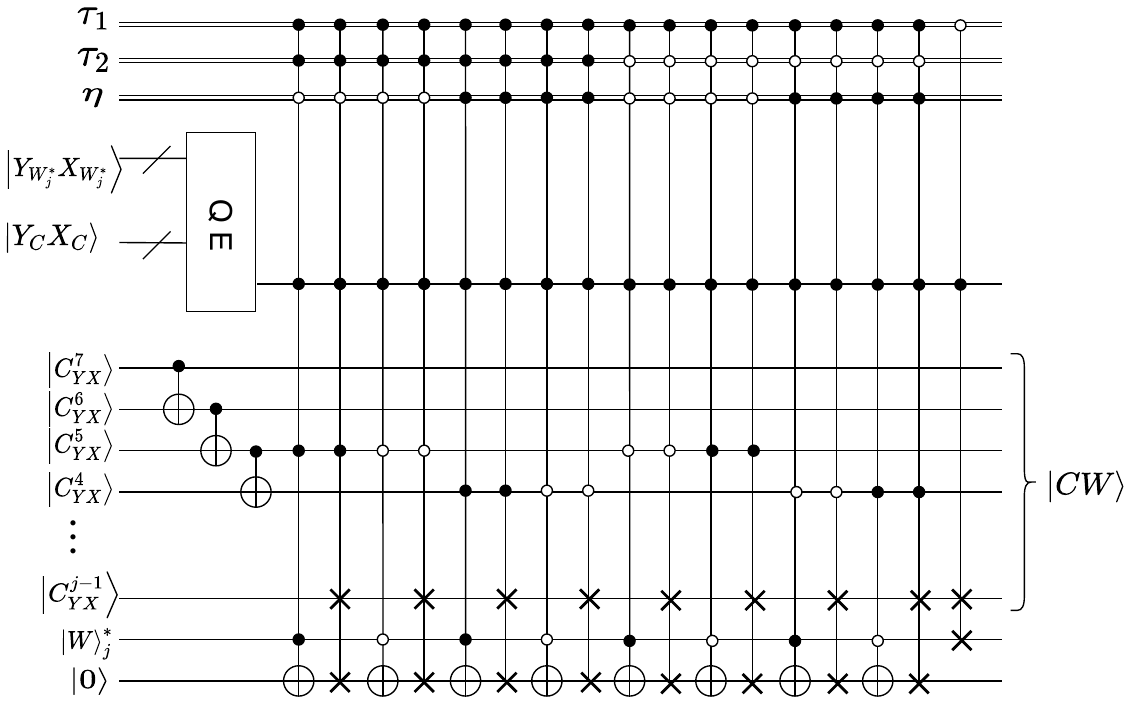}
	\caption{Quantum circuit for watermark embedding.}
	\label{FIG:cir_embed}
\end{figure}

After presenting the proposed HDWM, we next describe the elaborate embedding process. Based on the sizes of watermark and carrier images, the scale factor $r$ can be calculated as $n-m$. Figure \ref{FIG:embed} illustrates the flowchart of quantum watermark embedding. 
Preparation for quantum representation of classical images is necessary before embedding the watermark. The quantum watermark image $\ket{W}$ and the carrier image $\ket{C}$ are converted from the classical images using NEQR as represented by the Eqs.(\ref{eq4}), (\ref{eq5}).
\begin{equation}\label{eq4}
    \ket{W}=\frac{1}{2^{n-1} } \sum_{Y_{W} =0}^{2^{n-1}-1} \sum_{X_{W} =0}^{2^{n-1}-1}\left | W_{Y_WX_W} \right \rangle \left | Y_WX_W \right \rangle
\end{equation}
\vspace{0cm}
\begin{equation}\label{eq5}
    \ket{C}=\frac{1}{2^{n} } \sum_{Y=0}^{2^{n}-1} \sum_{X=0}^{2^{n}-1}\left | C_{YX} \right \rangle \left | YX \right \rangle
\end{equation}
where $\ket{Y_WX_W}$ and $\ket{YX}$ stand for the coordinate information, Note that $\ket{Y_WX_W}$ is indicated by entangled $n-1$ qubits sequences $\ket{Y_W}=\ket{y_{n-2}y_{n-3}...y_{i}...y_{0}}$ and $\ket{X_W}=\ket{x_{n-2}x_{n-3}...x_{i}...x_{0}}$, $\ket{x_{i}},\ket{y_{i}}\in \{\ket{0},\ket{1}\}$. Similarly, $\ket{YX}$ is expressed with two $n-1$ qubits sequences. $\ket{W_{Y_WX_W}}$ and $\ket{C_{YX}}$ indicate the pixel values, which are stated as $ \ket{W_{Y_WX_W}}=\otimes _{k=0}^{7} \ket{w_{Y_WX_W}^{k}}$, $    \ket{C_{YX}}=\otimes _{k=0}^{7} \ket{c_{YX}^{k}}$, where $\ket{w_{Y_WX_W}^{k}},\ket{c_{YX}^{k}}\in \{\ket{0},\ket{1}\}$, $k$ means the $k$-th qubit in the binary of pixel values. The symbol `$\otimes$' is tensor product.  After the preparation work is completed, subsequently, the quantum watermark embedding procedure, which consists of three steps, is described as below.
\begin{itemize}

    \item Step 1: The scale factor $r$ is calculated based on the carrier image and the watermark image, and the embedding parameters $\tau_1$ and $\tau_2$ are calculated depending on the histogram distribution of the watermark image. The scale factor $r$ is used for the AQSM in the 2nd step, and the embedding parameters $\tau_1$ and $\tau_2$ are applied to the watermark embedding in the 3rd step.

    \item Step 2: The AQSM is applied to the quantum watermark image processing to expand and transform it into a binary image. And the level $d$ of the AQSM is set as follows.
\begin{equation}
    d=\left\{
\begin{aligned}
&1, &  & r=1 \\
&r, &  & r>1 
\end{aligned}
\right.
\end{equation}
    
Specifically, the quantum watermark $\ket{W}$ and $r$ are used as inputs and obtain $\ket{W}_{j}^{\ast}(j=1,2,...,q)$ using AQSM. Note that when $r=1$, $q$ is 4 and when $r=2$, $q$ is 1. A special case needs to be illustrated here, when $r=1$, the output is four binary images with size $2^{n}\times2^{n}$, however, only three of the AQSM watermarks are actually embedded in view of the limited space. The three AQSM watermarks are got from the sets $\{\ket{w}_{1l}, \ket{w}_{2l},\ket{w}_{3l},\ket{w}_{4l}\}$, $\{\ket{w}_{5h}, \ket{w}_{7h},\ket{w}_{7h},\ket{w}_{7h}\}$, and $\{\ket{w}_{6h}, \ket{w}_{8h},\ket{w}_{8h},\ket{w}_{8h}\}$, respectively. Ultimately, the quantum watermark information to be embedded is noted as AQSM watermarks $\ket{W}_{j}^{\ast}$, $j\in \{1,2,3\}$.

\begin{table}[h]
\centering
\caption{Embedding rules of Histogram Distribution-based Watermarking Mechanism (HDWM).}\label{tbl1}
\begin{tabular}{rcccll }
\hline
\multicolumn{2}{l}{\textbf{Watermarking parameters}}&\multicolumn{4}{c}{\textbf{Embedding implementation with $V_{\eta}$}}\\ \hline
\multirow{2}{*}{$\tau_{1}$} & \multirow{2}{*}{$\tau_{2}$} & \multirow{2}{*}{$V_{\eta}$} & \multirow{2}{*}{$\ket{W}^{\ast}_{j}$} & \multicolumn{2}{c}{$i$-LSB ($i \in \{1,2,3\}$)} \\ \cline{5-6} 
 &  &  &  & $\ket{0}$ & $\ket{1}$ \\ \hline
\multirow{8}{*}{1} & \multirow{4}{*}{1} & \multirow{2}{*}{$\ket{0}$} & $\ket{0}$ & Flip & No change \\
 &  &  & $\ket{1}$ & No change & Flip \\ 
 &  & \multirow{2}{*}{$\ket{1}$} & $\ket{0}$ & No change & Flip \\
 &  &  & $\ket{1}$ & Flip & No change \\ 
 & \multirow{4}{*}{0} & \multirow{2}{*}{$\ket{0}$} & $\ket{0}$ & No change & Flip \\
 &  &  & $\ket{1}$ & Flip & No change \\ 
 &  & \multirow{2}{*}{$\ket{1}$} & $\ket{0}$ & Flip & No change \\
 &  &  & $\ket{1}$ & No change & Flip \\ 
\multirow{2}{*}{0} & \multirow{2}{*}{/} & \multirow{2}{*}{/} & $\ket{0}$ & No change & Flip \\
 &  &  & $\ket{1}$ & Flip & No change \\ \hline
\end{tabular}
\label{hdem_table}
\end{table}

\item Step 3: Finally, the quantum watermark information is embedded and the quantum watermarked image $\ket{CW}$ is obtained. The embedding procedure is accomplished using the embedding rules of HDWM, which are shown in the Table \ref{hdem_table}. Note that the diversion threshold $\lambda$ in the HDWM is set to 0.5. The XOR index $\eta$ is determined according to the scale factor $r$, along with the embedding parameters $\tau_1$ and $\tau_2$ to determine the embedding rules, and the AQSM watermarks $\ket{W}_{j}^{\ast}$, $j\in \{1,2,3\}$ are uniformly embedded in the quantum carrier image $\ket{C}$, the embedding starts from the first LSB of the carrier image in turn. Finally, the quantum watermarked image $\ket{CW}$ is obtained. When visualization is needed, only quantum measurements are required to turn the quantum watermarked image into a classical image. The embedded quantum circuit is shown in Figure \ref{FIG:cir_embed}.    
\end{itemize}
 
\subsection{Quantum watermark extraction using HDWM}
\begin{figure}[h]
	\centering
		\includegraphics[scale=.5]{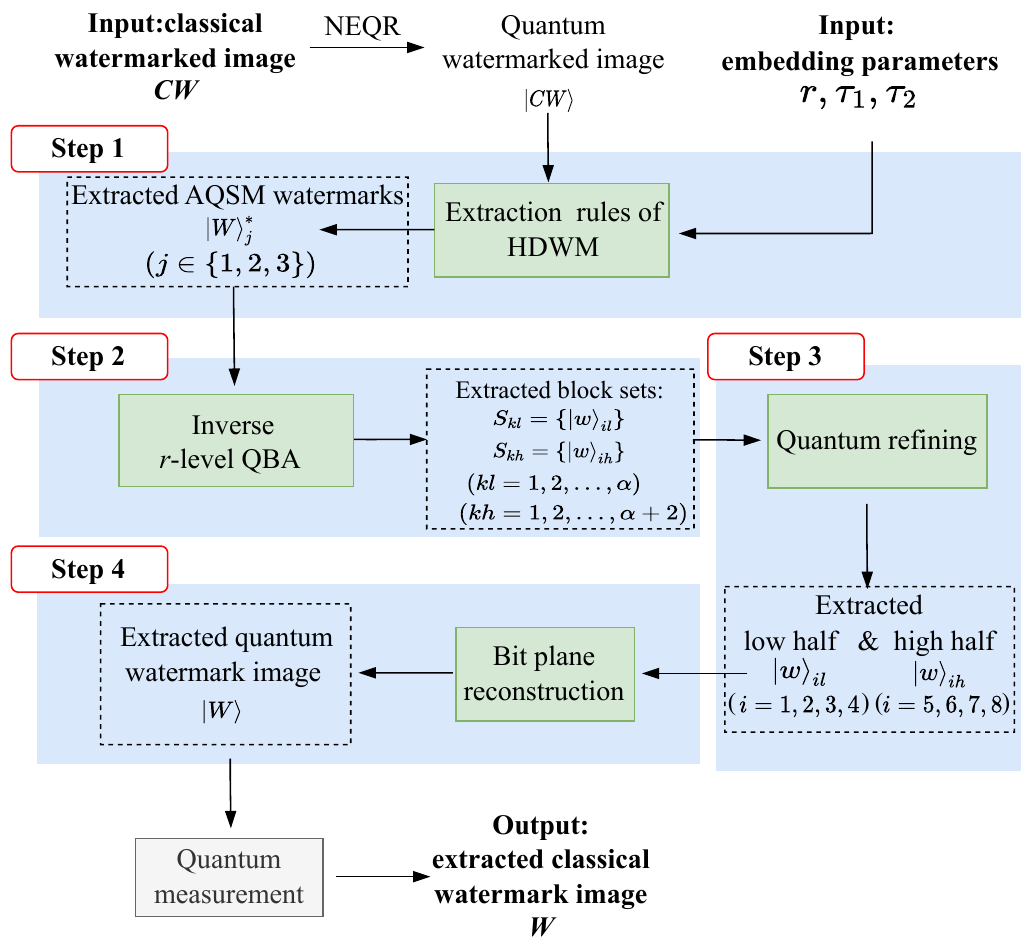}
	\caption{Flowchart of quantum watermark extraction.}
	\label{FIG:extract}
\end{figure}

Figure \ref{FIG:extract} illustrates the whole watermark extraction process of the proposed quantum watermarking scheme, which mainly includes watermark information extraction, inverse QBA module and quantum refining.
Before extracting the watermark, the NEQR method is applied to obtained the quantum watermarked image from the classical watermarked image. Then, $\ket{CW}$ is used for extraction along with the required parameters. Specifically, the extraction process consists of four steps, which are described in detail below.

\begin{itemize}
    \item Step 1: Corresponding to the embedding process, three empty quantum images $\ket{W}$ of size $2^{n}\times2^{n}$ with pixel values of $\ket{0}$ are prepared to store the extracted AQSM watermarks when the scale factor $r=1$, otherwise one is prepared. Similarly, the extraction rules are also formulated based on the proposed HDWM, and the parameters $\tau_1$ and $\tau_2$ determine the extraction strategy. Note that the AQSM watermarks $\ket{W}_{j}^{\ast}$, $j\in \{1,2,3\}$ are extracted sequentially starting from the first LSB of the watermarked image. The extraction rules of HDWM are shown in Table \ref{hdem_ex_table}. The corresponding watermark extraction quantum circuit is shown in Figure \ref{FIG:cir_extraction} (a).
\begin{table}[h]
\caption{Extraction rules of Histogram Distribution-based Watermarking Mechanism (HDWM).}\label{tbl1}
\begin{tabular}{rcccl }
\hline
\multicolumn{2}{l}{\textbf{Watermarking parameters}}&\multicolumn{3}{c}{\textbf{Extraction implementation with $V_{\eta}$}}\\ \hline
\multirow{1}{*}{$\tau_{1}$} & \multirow{1}{*}{$\tau_{2}$} & \multirow{1}{*}{$V_{\eta}$} & \multicolumn{1}{l}{$i$-LSB ($i \in \{1,2,3\}$)} & \multicolumn{1}{c}{$\ket{E_{j}}$} \\ \hline
\multirow{8}{*}{1} & \multirow{4}{*}{1} & \multirow{2}{*}{$\ket{0}$} & $\ket{0}$ & Flip  \\
 &  &  & $\ket{1}$ & No change \\ 
 &  & \multirow{2}{*}{$\ket{1}$} & $\ket{0}$ & No change  \\
 &  &  & $\ket{1}$ & Flip  \\ 
 & \multirow{4}{*}{0} & \multirow{2}{*}{$\ket{0}$} & $\ket{0}$ & No change  \\
 &  &  & $\ket{1}$ & Flip  \\ 
 &  & \multirow{2}{*}{$\ket{1}$} & $\ket{0}$ & Flip  \\
 &  &  & $\ket{1}$ & No change  \\ 
\multirow{2}{*}{0} & \multirow{2}{*}{/} & \multirow{2}{*}{/} & $\ket{0}$ & No change  \\
 &  &  & $\ket{1}$ & Flip  \\ \hline
\end{tabular}
\label{hdem_ex_table}
\end{table}

    \item Step 2: The inverse QBA module is applied to the extracted AQSM watermarks $\ket{W}_{j}^{\ast}$, $j\in \{1,2,3\}$, and then the extracted block sets, i.e. the high half block sets $S_{kl}={\ket{w}_{il}} (kl=1,2,...,\alpha)$ and the low half block sets $S_{kh}={\ket{w}_{ih}} (kl=1,2,...,\alpha+2)$ are obtained. It is worth emphasizing that all the watermark information is carried by the $2\alpha+2$ bit planes in these block sets.
\begin{figure*}[h]
	\centering
		\includegraphics[scale=.45]{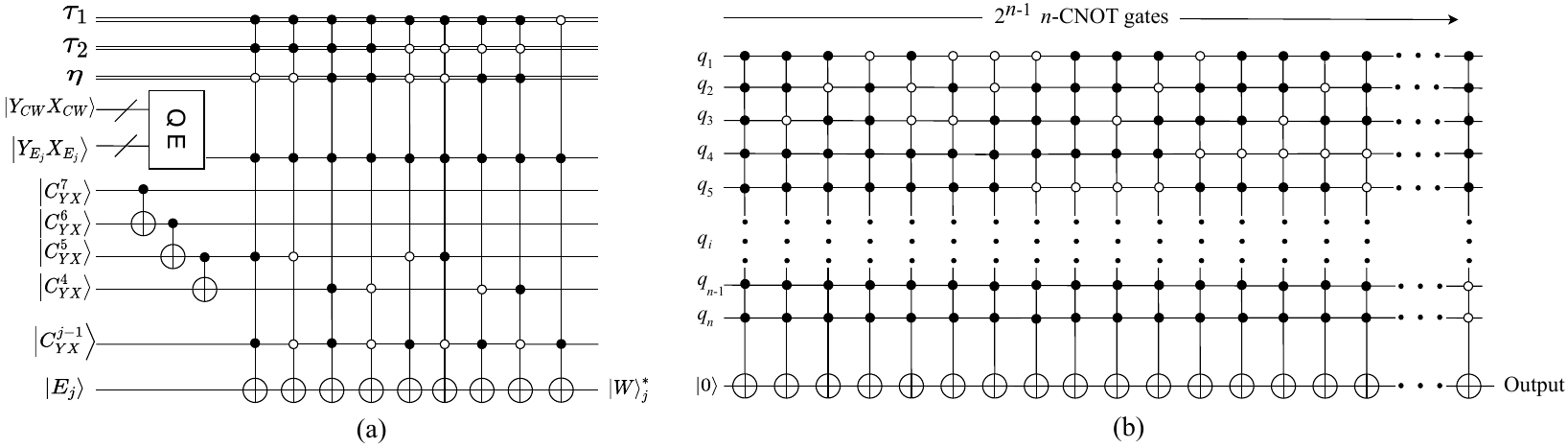}
	\caption{Quantum circuits for watermark extraction. (a) Circuit of watermark extraction,(b) Circuit of quantum refining.}
	\label{FIG:cir_extraction}
\end{figure*}
    
    \item Step 3: In order to improve the correct rate of extracting watermark information, the quantum refining method is proposed. Figure \ref{FIG:cir_extraction}(b) shows the circuit of quantum refining, and the core idea of quantum refining is to find the majority of the sequence of odd qubits as the logical information carried while ignoring the minority of sequence. By applying quantum refining approach, as a result, the low half $\ket{w}_{il}(i=1,2,3,4)$ and the high half $\ket{w}_{ih}(i=5,6,7,8)$ are extracted.
    \item Step 4: The extracted eight bit planes are recovered as the extracted quantum watermark $\ket{W}$ by bit plane reconstruction. It can be turned into a classical image by quantum measurements.
\end{itemize}

\section{Experiments and analysis}\label{part5}
Quantum computers are not yet widely available. Therefore, we simulated the proposed quantum watermarking method using MATLAB 2020b on a conventional computer with an Intel(R) Core(TM) i7 CPU 2.30GHz 16.00-GB RAM.MATLAB simplifies the representation and processing of huge matrices and vectors, whereas quantum images can be treated as matrices, and thus MATLAB is suitable for simulation. A total of 83 grayscale carrier images of size $512\times512$ from the dataset USC-SIPI\citep{weber2006usc} are used for the experiments in this paper. The watermark images used in the experiments are grayscale images of sizes $256\times256$, $128\times128$, and $64\times64$, respectively, `Elsevier', which is derived from the logo of Elsevier Publishing. To demonstrate the effectiveness of our proposed scheme, we used Peak Signal-to-Noise Ratio (\emph{PSNR}), Structural Similarity Index Measure (\emph{SSIM}), and Normalized Cross-Correlation (\emph{NCC}) metrics \citep{koley2022feature} to evaluate the experimental results. 
\subsection{Capacity analysis}
In quantum watermarking schemes, the embedding capacity is also a metric to evaluate the efficiency of the watermarking algorithm. In this paper, the embedding capacity is denoted as $C$ which is the ratio of the total number of qubits $Q_w$ in the embedded watermark image to the total number of qubits $Q_c$ in the carrier image. For the proposed quantum watermarking scheme, the sizes of watermark image and carrier image are $2^{n-r}\times2^{n-r}$ and $2^{n}\times2^{n}$, respectively. Thus, the embedding capacity depends on the scale factor $r$, and the embedding capacity is calculated as follows.
\begin{equation}
    C=\frac{\textit{qubits of watermark image $Q_w$}}{\textit{qubits of carrier image $Q_c$}}=\frac{2^{n-r}\times2^{n-r}\times8}{2^{n}\times2^{n}\times 8}=\frac{1}{4^{r}}
\end{equation}
In summary, it is not difficult to calculate the embedding capacity with the three scale factors, i.e. the embedding capacity is 1/4 when $r=1$, 1/16 as $r=2$, and 1/64 with $r=3$, respectively.
\begin{figure*}[h]
\centering		
\includegraphics[scale=.53]{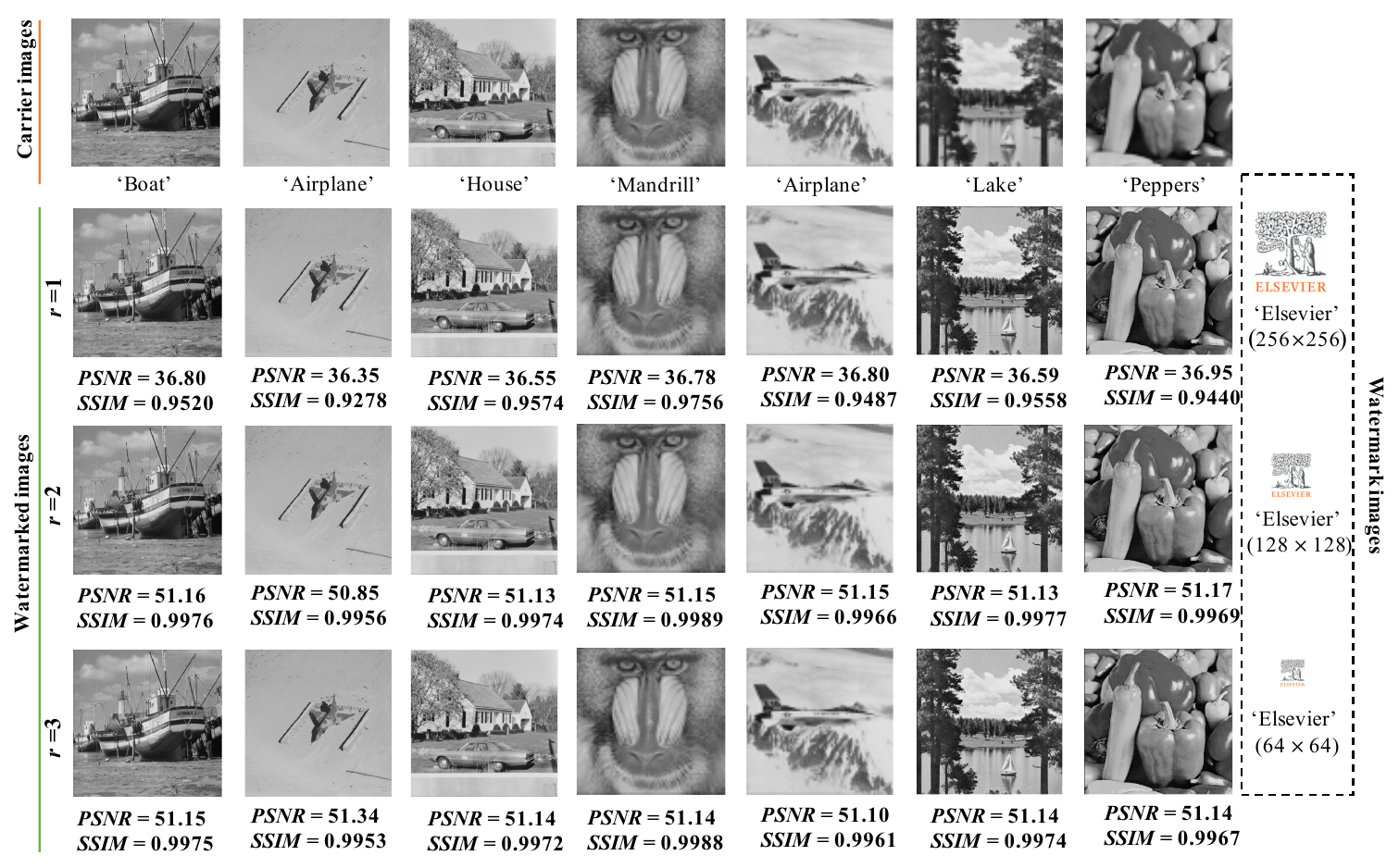}
\caption{Demonstration of quality of watermarked images. 1st row: carrier images, 2nd row: corresponding watermarked images when $r=1$ and $256\times256$ sized watermark `Elsevier', 3rd row: corresponding watermarked images when $r=2$ and $128\times128$ sized watermark `Elsevier', 4rd row: corresponding watermarked images when $r=3$ and $64\times64$ sized watermark `Elsevier'.}
\label{FIG:visual_demo}
\end{figure*}

\subsection{Visual quality analysis}
Visual quality assessment refers to the quality evaluation of a digital image and giving specific numerical metrics. Common visual quality evaluation metrics include PSNR, SSIM, etc. These metrics are usually based on the cognitive characteristics of the human visual system to evaluate the quality of an image by comparing the original image with the processed image. Figure \ref{FIG:visual_demo} gives a comparison of seven randomly selected carrier images and their corresponding watermarked images, providing \emph{PSNR} and \emph{SSIM} values. The 1st row is carrier images, the 2nd row is corresponding watermarked images when $r=1$ and $256\times256$ sized watermark `Elsevier', the 3rd row is corresponding watermarked images when $r=2$ and $128\times128$ sized watermark `Elsevier', and the 4rd row is corresponding watermarked images when $r=3$ and $64\times64$ sized watermark `Elsevier'. As is obvious, visually, there is no significant difference between them. 

\begin{figure}[h]
	\centering
 \subfigure[]{
		\includegraphics[scale=.11]{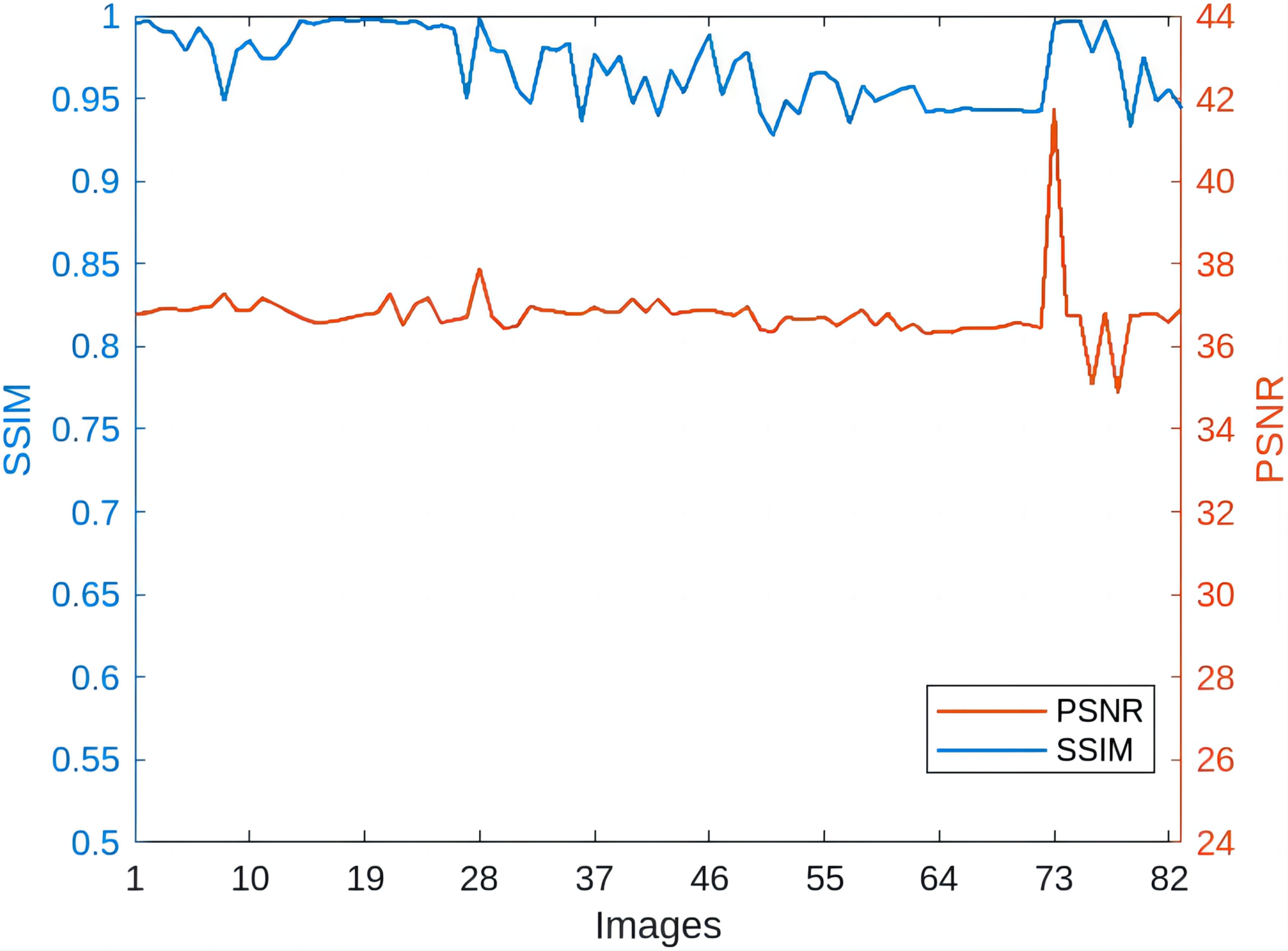}}
  \subfigure[]{
		\includegraphics[scale=.11]{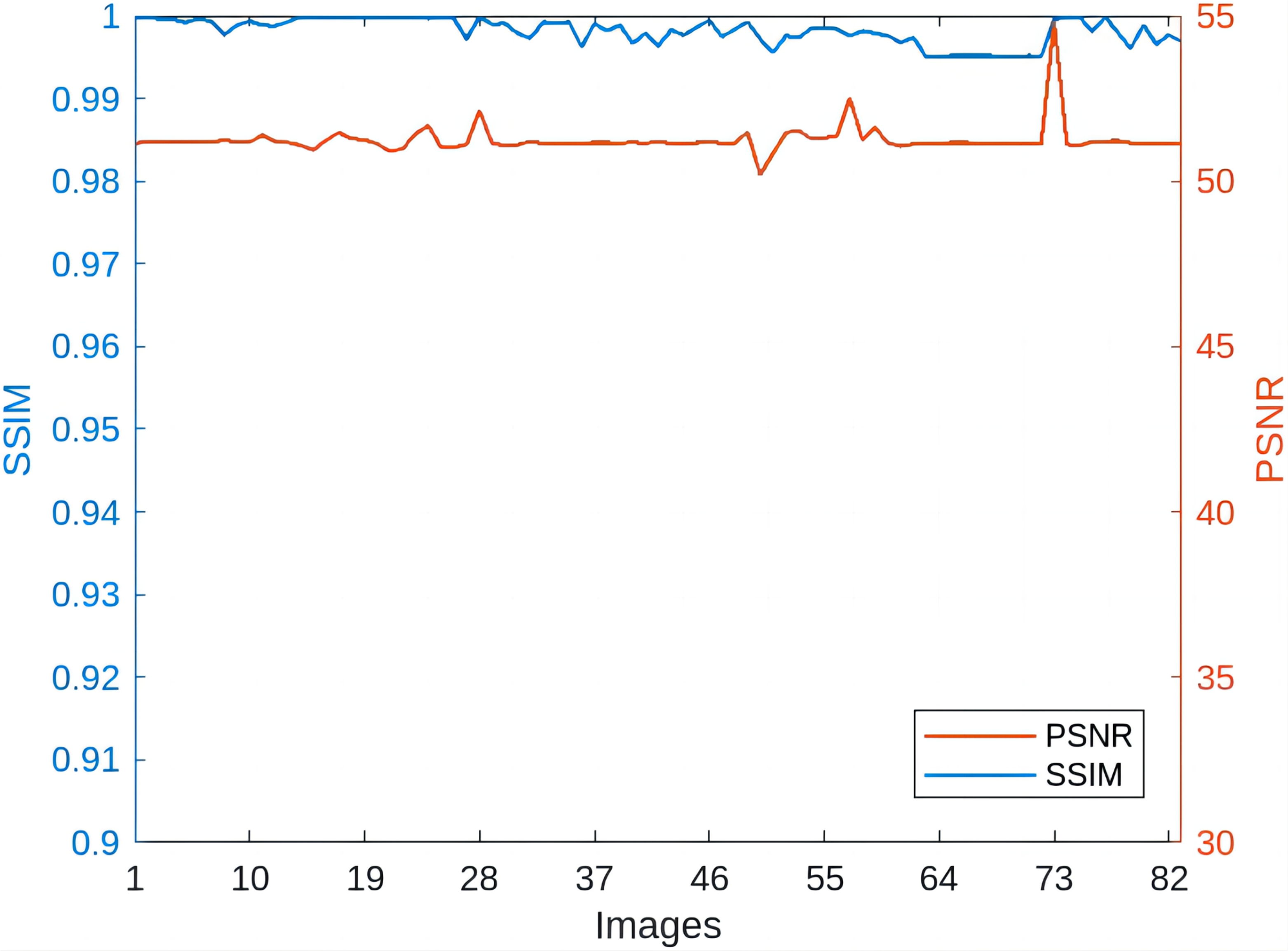}}
  \subfigure[]{
		\includegraphics[scale=.113]{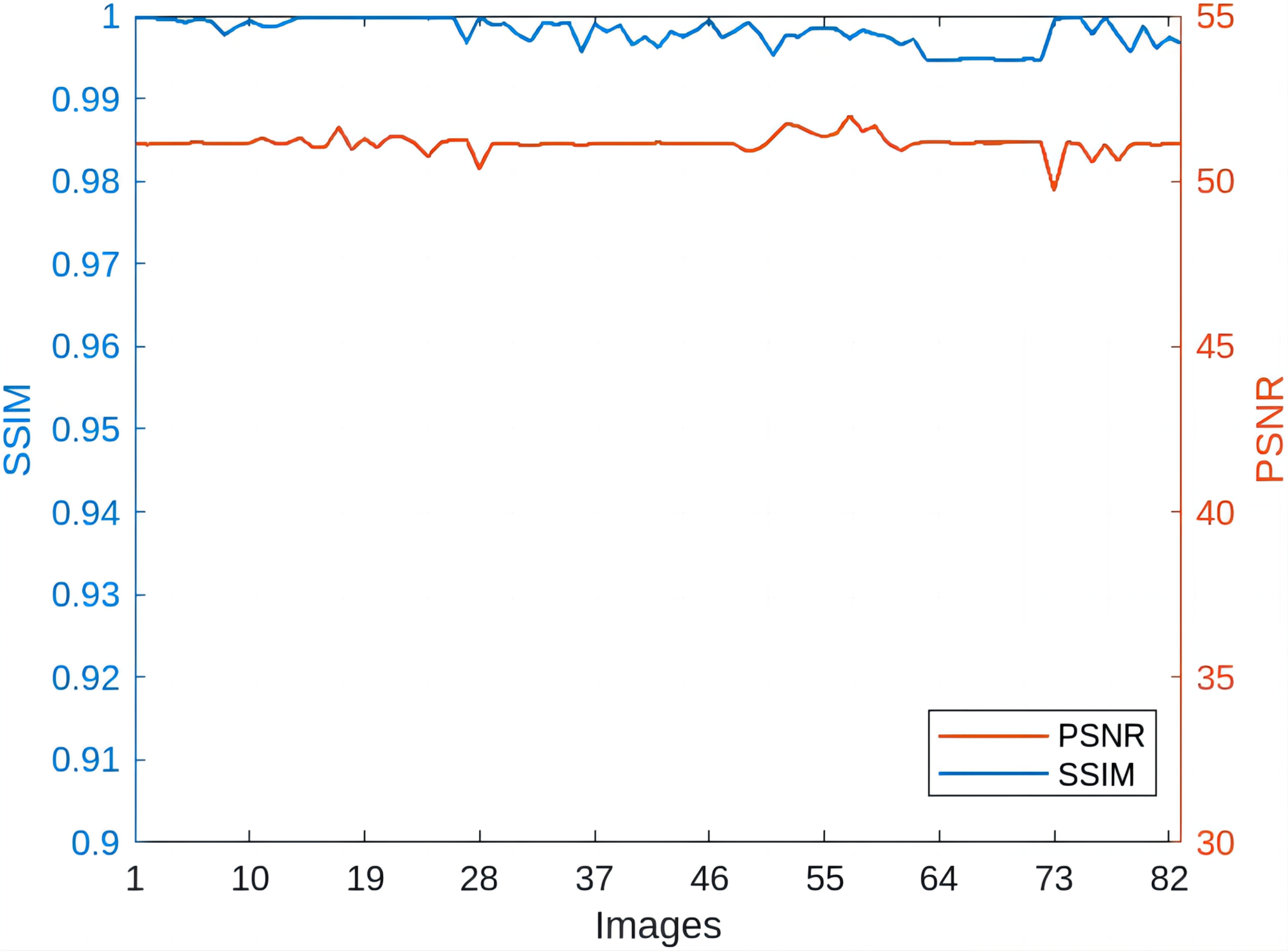}}
\caption{The visual quality of the watermarked images in terms of \emph{SSIM} and \emph{PSNR}. (a) Results of $r=1$, (b) results of $r=2$, (b) results of $r=3$.}
	\label{FIG:visual_line}
\end{figure}
Figure \ref{FIG:visual_line} shows the \emph{PSNR} and \emph{SSIM} values of the watermarked images, where the (a) shows the results for an embedding capacity of 1/4 ($r=1$), the (b) shows the results for an embedding capacity of 1/16 ($r=2$), and the (c) is the results for an embedding capacity of 1/64 ($r=3$). From the results, it can be concluded that the mean values of \emph{PSNR} for the three embedding capacities are 36.78 dB, 51.25 dB and 51.15 dB, respectively, while the mean values of \emph{SSIM} is 0.9693, 0.9982 and 0.9981, which are very close to 1. Overall, the visual effect of the carrier images is not disturbed by the embedding of the watermark, and from the results, the \emph{PSNR} and \emph{SSIM} values of the watermarked images indicate that the image quality is not significantly affected. Therefore the proposed quantum watermark embedding is imperceptible.


\subsection{Robustness analysis}
\begin{figure}[h]
	\centering		\includegraphics[scale=.3]{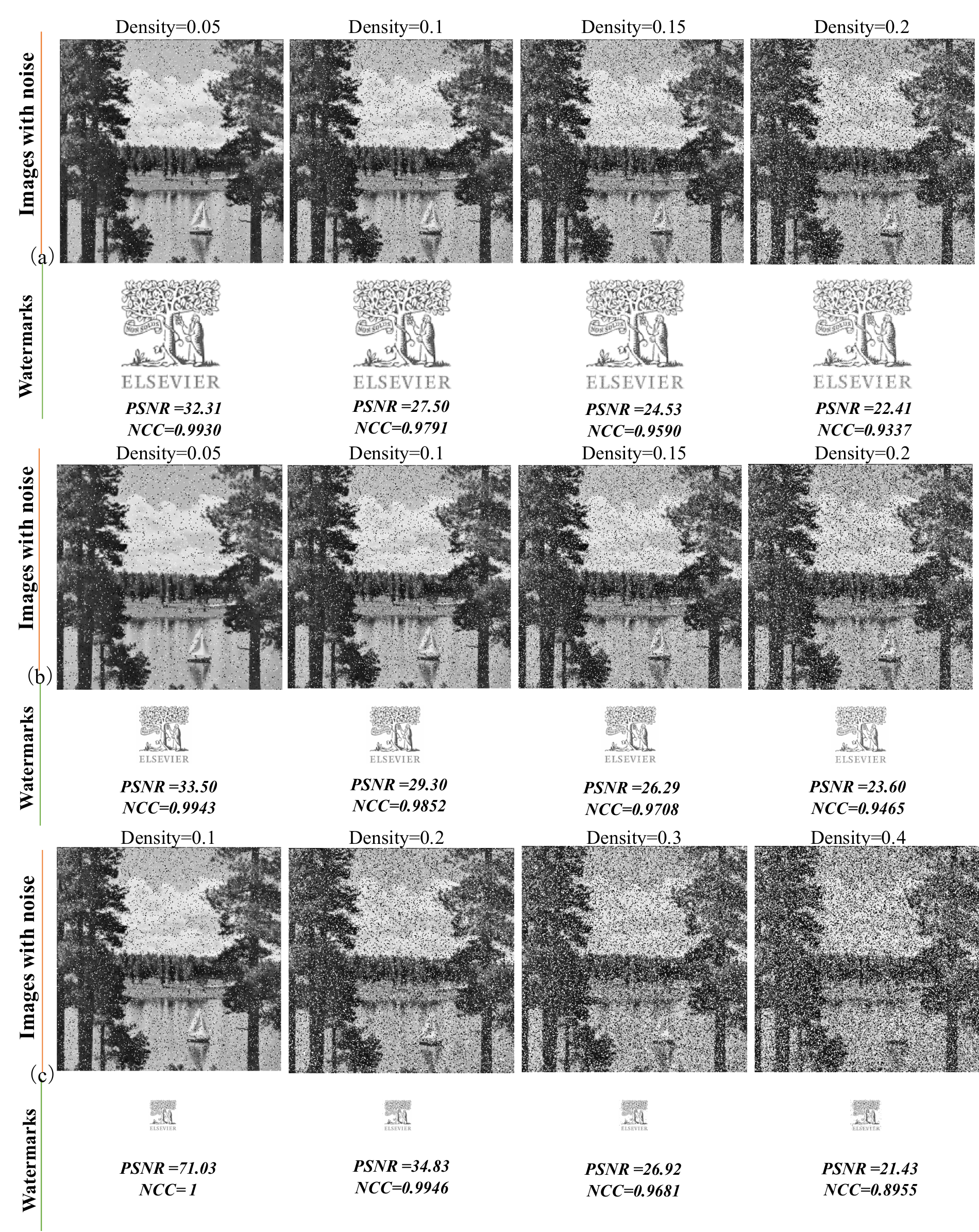}
	\caption{Visualization of robustness against `Salt and Pepper' noise addition. (a) The demonstration of $r=1$ with the noise density of 0.05, 0.1, 0.15 and 0.2, (b) The demonstration of $r=2$ with the noise density of 0.05, 0.1, 0.15 and 0.2, (c) The demonstration of $r=3$ with the noise density of 0.1, 0.2, 0.3 and 0.4.}
	\label{FIG:noise_demo}
\end{figure}

\begin{figure}[h]
	\centering
 \subfigure[]{
		\includegraphics[scale=.11]{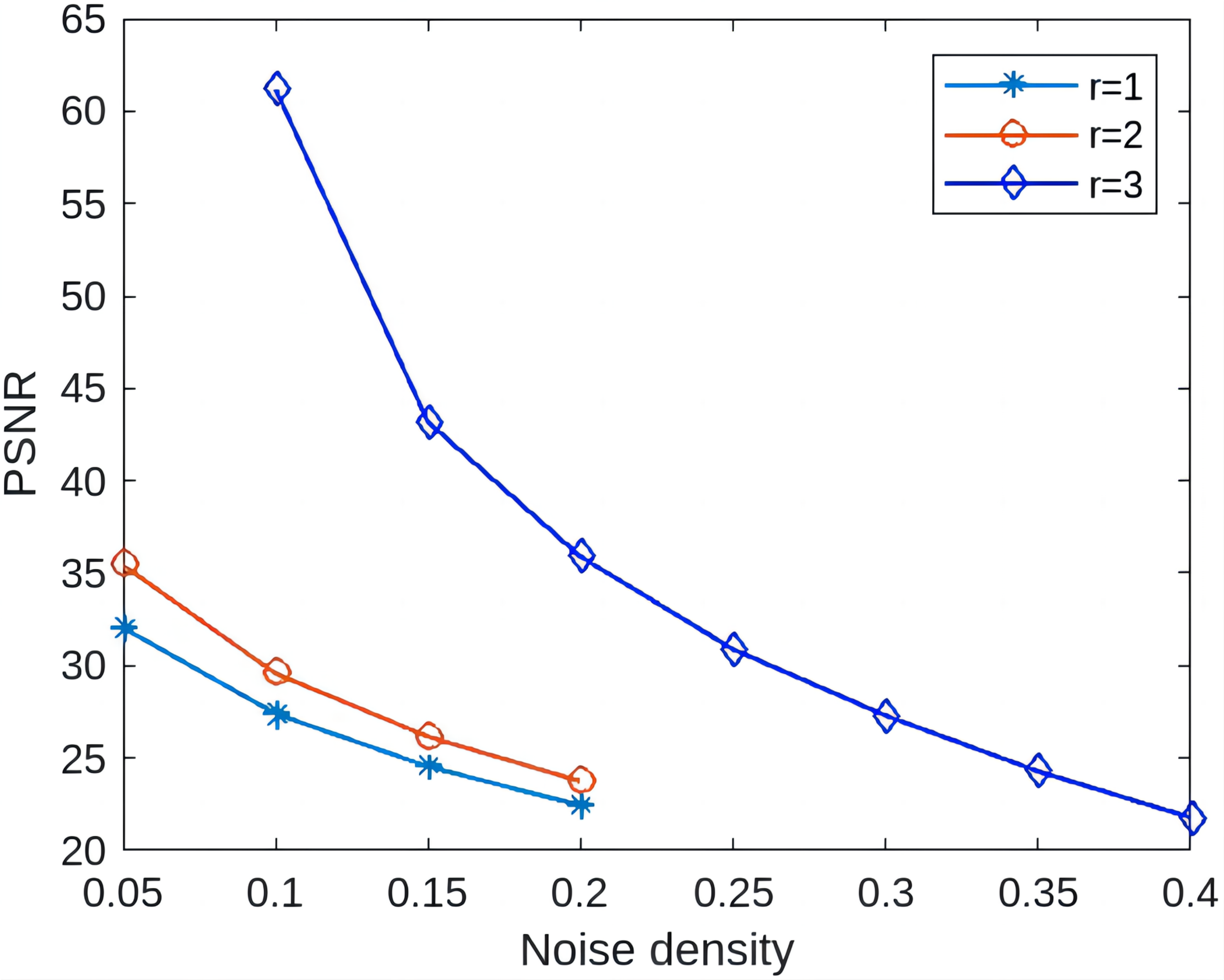}}
  \subfigure[]{
		\includegraphics[scale=.16]{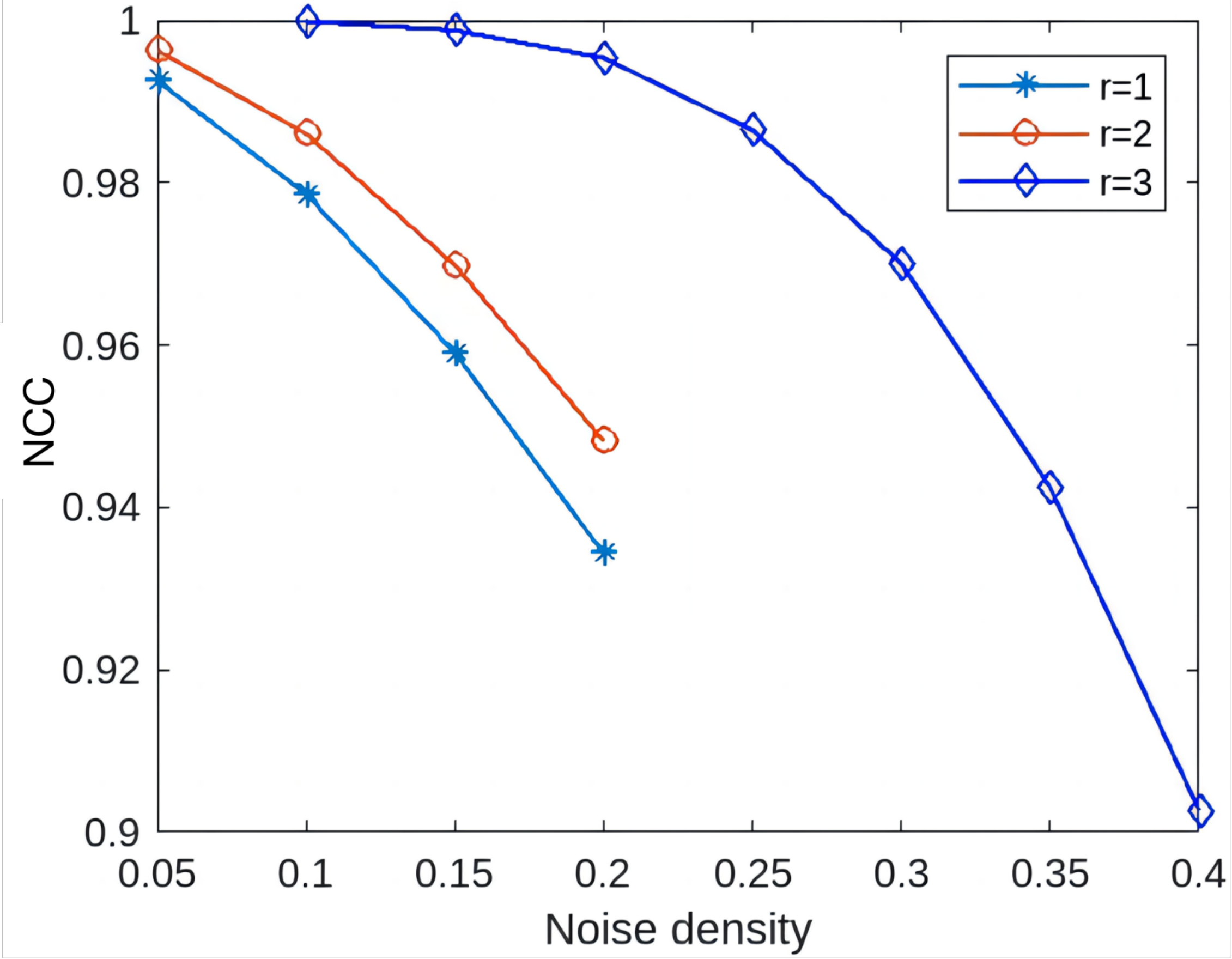}}
	\caption{The robustness of proposed scheme under `Salt and Pepper' noises addition in terms of (a) average \emph{PSNR} values, (b) average \emph{NCC} values.}
	\label{FIG:noise_line}
\end{figure}

To evaluate the robustness of the proposed quantum watermarking method, we attack the watermarked images with `Salt and Pepper' noise and cropping. The \emph{PSNR} and \emph{NCC} values are used to evaluate the quality of the watermarked images extracted from the corrupted watermarked images. It should be noted that we only show the results for \emph{PSNR} values above 20 dB, so the noise density is within 0.2 when the embedding capacity is 1/4 ($r=1$) and 1/16 ($r=2$), and within 0.4 when the embedding capacity is and 1/64 ($r=3$). In particular, the `Salt and pepper' noise with densities of 0.05, 0.1, 0.15, and 0.2 are added to the watermarked images when the embedding capacity was 1/4 ($r=1$) and 1/16 ($r=2$), while the `Salt and pepper' noise with densities of 0.1, 0.2, 0.3, and 0.4 are added to the watermarked images when the embedding capacity was 1/64 ($r=3$). Figure \ref{FIG:noise_demo} shows the demonstrations of the three embedding capacities, where (a) is the demonstration of $r=1$ with the noise density of 0.05, 0.1, 0.15 and 0.2, (b) is the demonstration of $r=2$ with the noise density of 0.05, 0.1, 0.15 and 0.2, and (c) is the demonstration of $r=3$ with the noise density of 0.1, 0.2, 0.3 and 0.4.
Figure \ref{FIG:noise_line} shows the test results for scale factors of $r=1$, $r=2$ and $r=3$, where (a) is the average \emph{PSNR} result, while (b) is the average \emph{NCC} result. At a noise density of 0.05, the average \emph{PSNR} values for embedding capacity of 1/4 ($r=1$) and 1/16 ($r=2$) are 32.06 dB and 35.45 dB, respectively, while the average \emph{NCC} values are 0.9926 and 0.9963. And at the noise density of 0.2, their average \emph{PSNR} values are 22.44 dB and 23.74 dB, respectively, while the average \emph{NCC} values are 0.9343 and 0.9481. For the embedding capacity of 1/64 ($r=3$), the average \emph{PSNR} and \emph{NCC} values are 61.20dB and 0.9999 for a noise density of 0.1, while the average \emph{PSNR} and \emph{NCC} values are 21.72dB and 0.9014 for a noise density of 0.4.

\begin{figure}[h]
	\centering
\includegraphics[scale=.33]{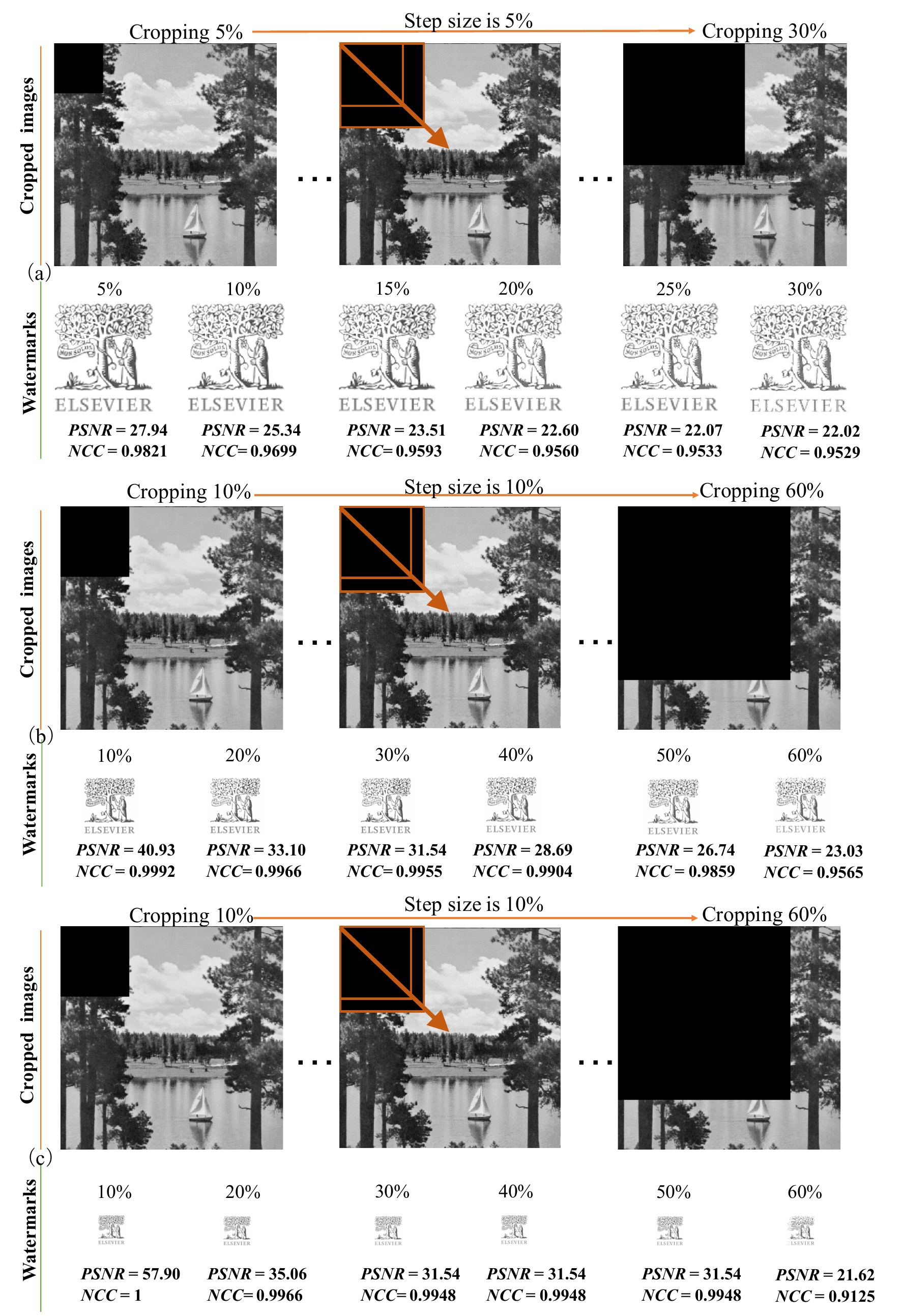}
	\caption{Visualization of robustness against cropping attack. (a) $r=1$ with cropping percentage varies from $5\%$ to $30\%$, (b) $r=2$ with cropping percentage varies from $10\%$ to $60\%$, (c) $r=3$ with cropping percentage varies from $10\%$ to $60\%$.}
	\label{FIG:crop_demo}
\end{figure}

\begin{figure}[h]
	\centering
 \subfigure[]{
		\includegraphics[scale=.11]{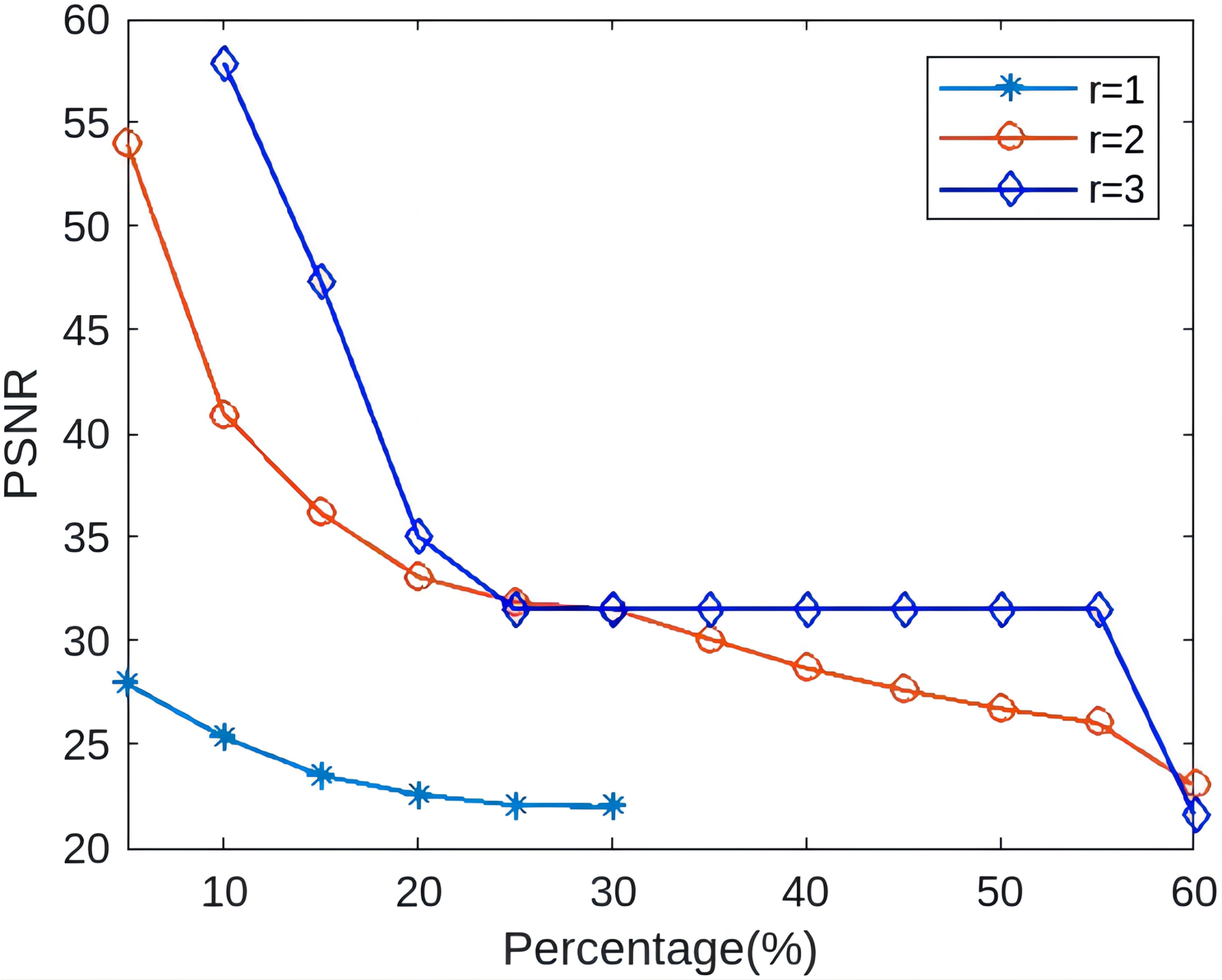}}
  \subfigure[]{
		\includegraphics[scale=.115]{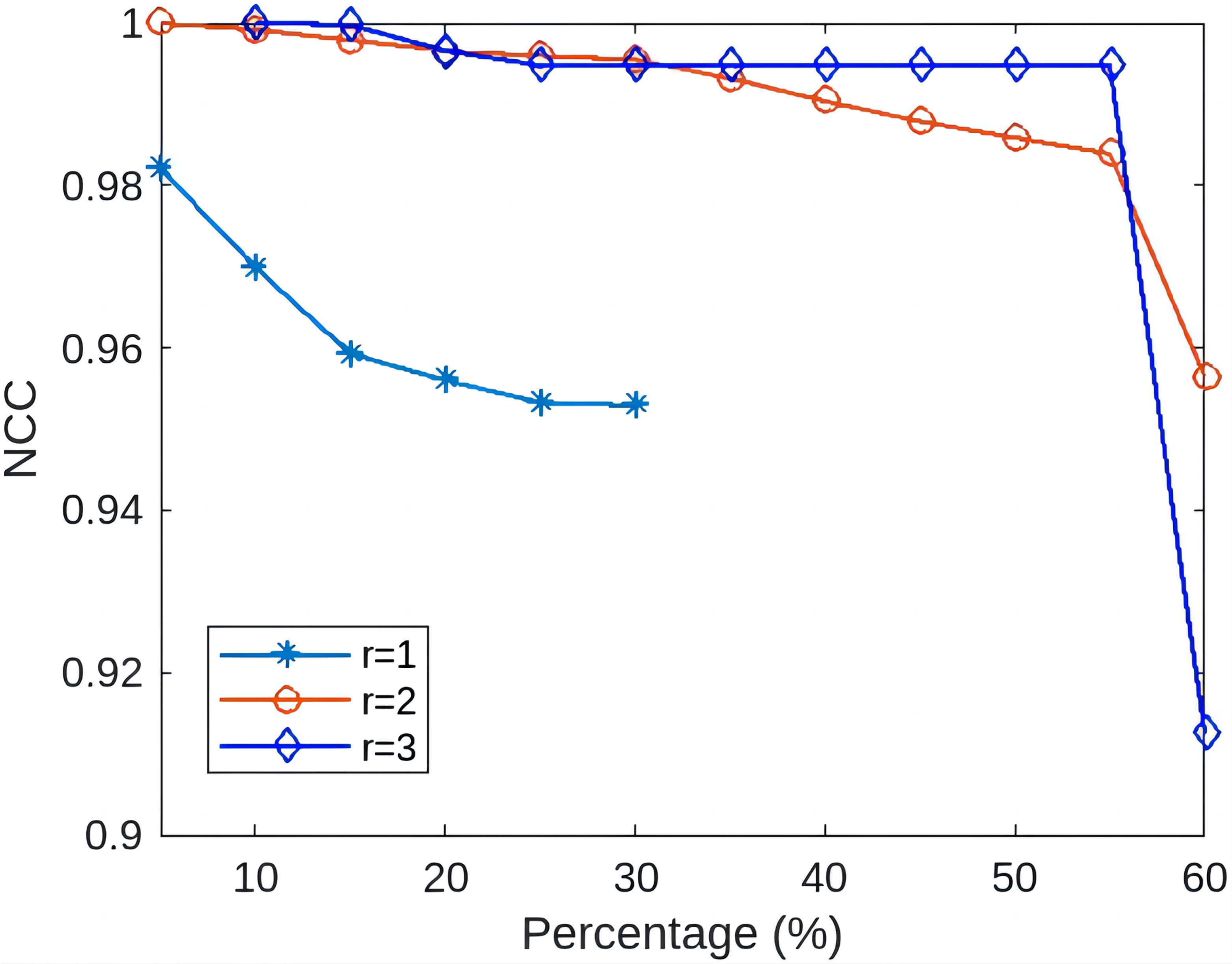}}
	\caption{The robustness of proposed scheme under cropping attack in terms of (a) average \emph{PSNR} values, (b) average \emph{NCC} values. }
	\label{FIG:crop_line}
\end{figure}

Regarding the cropping attack, we utilize a cropping method that starts from the top left vertex and expands the cropping square along the diagonal direction. Specifically, when the embedding capacity is 1/4 (r=1), the crop size starts at $5\%$, then 10\%, followed by increments of 5\% up to $30\%$. When the embedding capacity is 1/16 (r=2) and 1/64 (r=3), the cropping size starts at $5\%$, then $10\%$, followed by $5\%$ increments up to $60\%$. It is worth emphasizing that we only kept the results with \emph{PSNR} above 20 dB, so the cropping size is within 30\% when the embedding capacity is 1/4 ($r=1$). When the embedding capacity is 1/16 ($r=2$) and 1/64 ($r=3$), the cropping size is within 60\%.
Figure\ref{FIG:crop_demo} shows the Visualization of robustness against cropping attacks. The (a) is result of $r=1$ with cropping percentage varies from $5\%$ to $30\%$, and (b) is the result of $r=2$ with cropping percentage varies from $10\%$ to $60\%$, and (c) is the result of $r=3$ with cropping percentage varies from $10\%$ to $60\%$. Figure\ref{FIG:crop_line} shows the robustness of proposed scheme under cropping attack in terms of average \emph{PSNR} and \emph{NCC} values, where (a) is the average \emph{PSNR} values of $r=1,2,3$, while the (b) is the average \emph{NCC} values. Note that when the embedding capacity is 1/64 ($r=3$), the PSNR values of the extracted watermark images are almost `Inf' for only 5\% cropping area, so our result data is shown from 10\% cropping for $r=3$. From the results, it can be seen that when $r=1$, The average \emph{PSNR} values of the watermark extracted from the watermarked image with cropping 5\% is 27.94 dB and the average \emph{NCC} values is 0.9821, while the average \emph{PSNR} and \emph{NCC} values are 22.02 dB and 0.9529, respectively, when cropping 30\%. When $r=2$, the average \emph{PSNR} of the watermark extracted from the watermarked image with 5\% cropped is 54.00 dB and the average \emph{NCC} is 1, while the average \emph{PSNR} and \emph{NCC} values remain at 23.03 dB and 0.9565 even when they are cropped by 60\%. When r=3, the average \emph{PSNR} of the watermark extracted from the cropped 10\% watermarked images is 57.90dB and the average \emph{NCC} is 1. And the average \emph{PSNR} and \emph{NCC} values remain above 31.54dB and 0.9948 when the cropped area is within 55\%. Even when cropped by 60\%, the average \emph{PSNR} and \emph{NCC} values remain at 21.61 dB and 0.9125. 

It is clear that the AQSM-based quantum watermarking scheme reflects the imperceptibility of the embedded watermark at different sizes. In terms of robustness, all of them show good resistance to noise and cropping attacks. Even when the size factor $r$ is larger than 2, the watermarked images with high density noise addition and cropped by large area can still extract recognizable watermark images.

\subsection{Comparisons and analysis }
\begin{figure}[h]
	\centering
 \subfigure[]{
		\includegraphics[scale=.11]{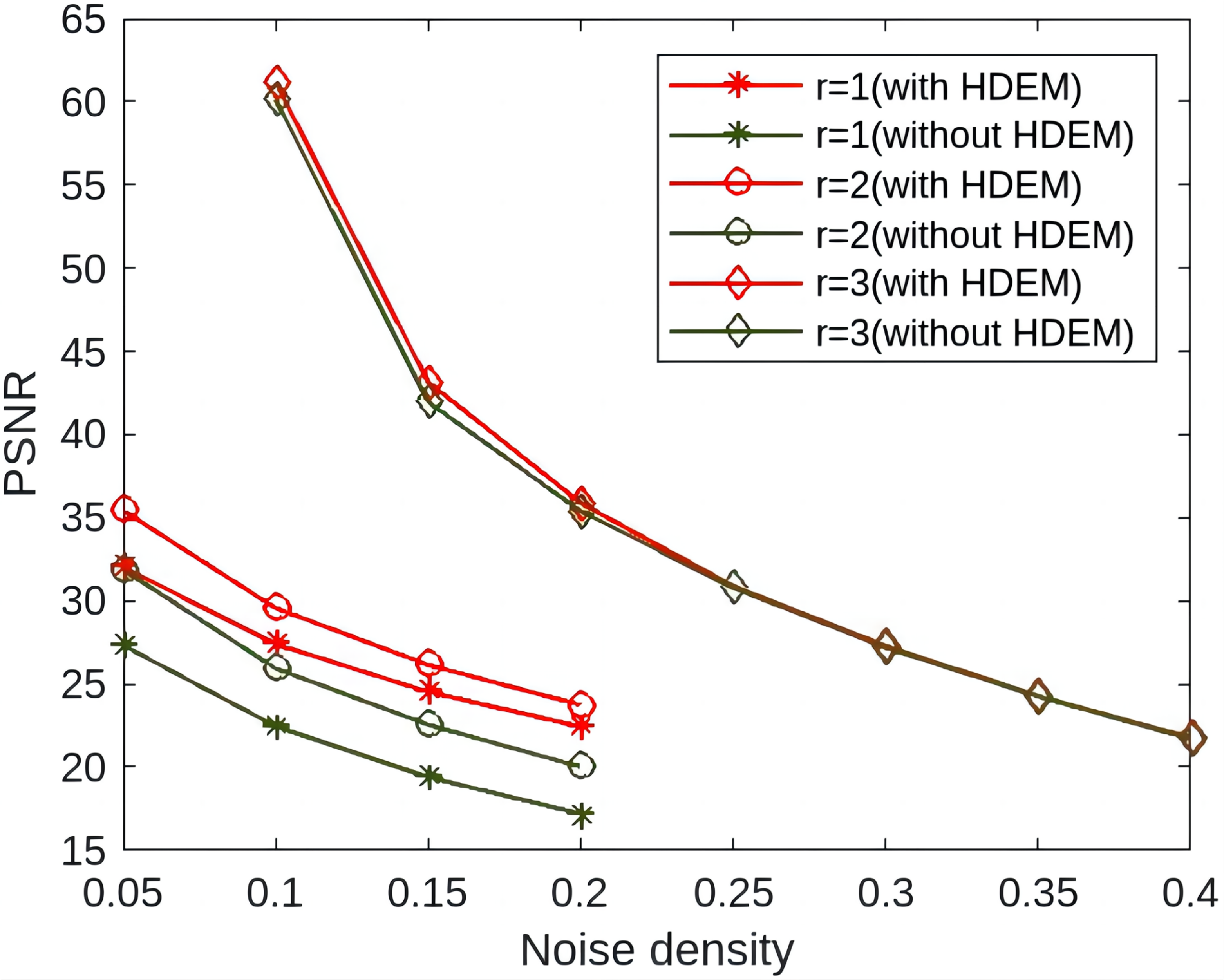}}
  \subfigure[]{
		\includegraphics[scale=.11]{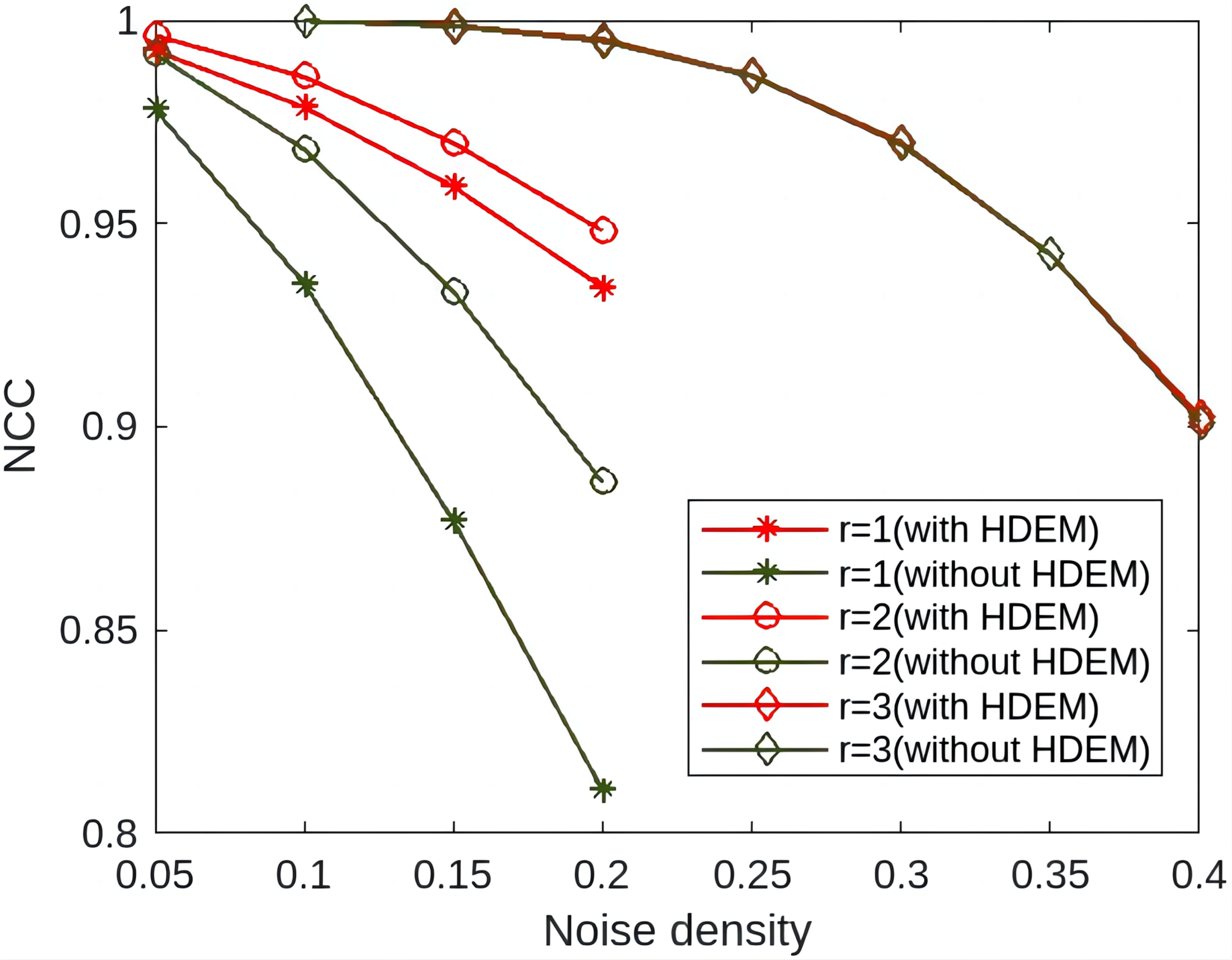}}
	\caption{Comparison of the results under `Salt and Pepper' noise addition with and without the HDWM in terms of (a) average \emph{PSNR} values, (b) average \emph{NCC} values. }
	\label{FIG:HDEM_COMP}
\end{figure}

To demonstrate the advantages of our proposed HDWM method in terms of robustness, i.e., improved resistance to `Salt and Pepper' noise, we compare the results with the HDWM method and those without HDWM but direct LSB substitution. Figure \ref{FIG:HDEM_COMP} shows the comparisons of the results with and without the HDWM method on the `Salt and Pepper' noise addition, where (a) is the comparison result of the average \emph{PSNR} values, while (b) is the comparison result of the average \emph{NCC} values. The red curve indicates the results with the HDWM method, while the green curve indicates the results without HDWM. It is obvious that the red curve is higher than the green curve in terms of the average \emph{PSNR} values and the average \emph{NCC} values. Quantitatively, the average \emph{PSNR} values of the results with HDWM improved by 23.14\%, 14.42\%, and 1.25\% over those without HSWM for the embedded capacities 1/4 $(r=1)$, 1/16 $(r=2)$, and 1/64 $(r=3)$, respectively, which illustrates the positive effect of the proposed HDWM method in improving the robustness.

To illustrate the value of our work, comparisons are made with other methods. Since the current quantum watermarking schemes are not diverse sufficient, with the same embedding capacity and image type, we only compare the results with an embedding capacity of 1/4 $(r=1)$ to the existing schemes, including the visual quality of watermarked images and the robustness of `Salt and Pepper' noise addition. The other two embedding capacities are not compared due to the lack of fair and reasonable comparators. Table \ref{comp_visual} shows the comparison results with existing works in terms of quality of watermarked images, where the bolded is the best result, the underlined is the second best result, and the italic is our result. As can be seen from the results, in terms of visual quality, the average of our results is 38.97 dB, which is only 4.13\% lower than the average of the best results from \citep{zeng2021quantum}. Even so, the embedded watermark does not affect the visual quality of the carrier images, and the slight decrease in the \emph{PSNR} of the visual quality is rewarded with a significant increase in robustness.

\begin{table}[h]\label{comp_visual}
\centering
\caption{Comparison with existing works in terms of quality of watermarked images. Bolded is the best result, underlined is the second best result, and italic is our result.} \label{compare_psnr}
\begin{tabular}{lccc }
\hline
\multirow{2}{*}{Carrier images} & \multicolumn{3}{c}{\begin{tabular}[c]{@{}c@{}}Methods (Capacity=1/4)\\ \emph{PSNR}/dB\end{tabular}} \\ \cline{2-4} 
                                & \multicolumn{1}{c}{ \cite{luo2019adaptive}}& \multicolumn{1}{c}{\cite{zeng2021quantum}}      & Proposed      \\ \hline
Barbara       & \multicolumn{1}{c}{34.27}               & \multicolumn{1}{c}{{\textbf{39.67}}}               & \emph{{\ul{38.90}}}      \\ 
Boat                            & \multicolumn{1}{c}{36.62}               &\multicolumn{1}{c}{{\textbf{42.00}}}                & \emph{{\ul{38.88}}}       \\ 
Bridge                          & \multicolumn{1}{c}{35.05}               & \multicolumn{1}{c}{{\textbf{40.46}}}                & \emph{{\ul{39.41}}}     \\ 
Couple                          & \multicolumn{1}{c}{34.44}               & \multicolumn{1}{c}{{\textbf{39.78}}}               & \emph{{\ul{38.92}}}    \\ 
Goldhill                        & \multicolumn{1}{c}{34.11}               & \multicolumn{1}{c}{{\textbf{39.40}}}               & \emph{{\ul{38.89}}}      \\ 
House                           & \multicolumn{1}{c}{36.25}               & \multicolumn{1}{c}{{\textbf{41.60}}}                & \emph{{\ul{38.87}}}       \\ 
Lake                            & \multicolumn{1}{c}{34.52}               & \multicolumn{1}{c}{{\textbf{40.00}}}                & \emph{{\ul{38.86}}}      \\ 
Plane                           & \multicolumn{1}{c}{37.01}               & \multicolumn{1}{c}{{\textbf{42.28}}}                & \emph{{\ul{39.04}}}       \\ \hline
\end{tabular}
\label{comp_visual}
\end{table}

In terms of robustness, the results of experiments on `Salt and Pepper' noise addition are compared with existing work. The noise with densities of 0.05 and 0.1 are added, respectively, and the corresponding \emph{PSNR} values of the extracted watermark images are calculated. Table \ref{comp_noise} shows the comparison results, where italicized is our result, and bolded is the best result, while underlined is the second best result. It is clear that our results are the best. Specifically, at a noise density of 0.05, the mean \emph{PSNR} values of our scheme is 21.28\% and 16.33\% higher than the mean \emph{PSNR} values of methods in \citep{luo2019adaptive} and \citep{zeng2021quantum}, respectively. Likewise, at a noise density of 0.1, the mean PSNR values of our scheme improved by 15.14\% and 10.22\% compared to those of \citep{luo2019adaptive} and \citep{zeng2021quantum}, respectively.

\begin{table*}[h]
\caption{Comparison with existing works on robustness against `Salt and Pepper' noise addition. Noises with density of 0.05 and 0.1 are respectively added, and the corresponding \emph{PSNR}/dB of the extracted watermarks are calculated. 
} 
\centering
\begin{tabular}{clcccccccccc}
\hline
\multicolumn{2}{l}{Carrier images} & \multicolumn{2}{c}{Barbara} & \multicolumn{2}{c}{Boat} & \multicolumn{2}{c}{Couple} & \multicolumn{2}{c}{Coco} & \multicolumn{2}{c}{Crowd} \\ \hline
\multicolumn{2}{l}{Watermark images} & \multicolumn{1}{c}{Zelda} & Plane & \multicolumn{1}{c}{Zelda} & Plane & \multicolumn{1}{c}{Zelda} & Plane & \multicolumn{1}{c}{Zelda} & Plane & \multicolumn{1}{c}{Zelda} & Plane \\ \hline
\multicolumn{1}{c}{\multirow{3}{*}{Density is 0.05}} & \cite{luo2019adaptive} & \multicolumn{1}{c}{22.81} & 24.51 & \multicolumn{1}{c}{22.51} & 24.54 & \multicolumn{1}{c}{22.63} & 24.61 & \multicolumn{1}{c}{22.30} & 24.09 & \multicolumn{1}{c}{22.31} & 24.69 \\ 
\multicolumn{1}{c}{} & \cite{zeng2021quantum} & \multicolumn{1}{c}{{\ul{23.52}}} & {\ul{25.25}} & \multicolumn{1}{c}{{\ul{23.85}}} & {\ul{25.00}} & \multicolumn{1}{c}{{\ul{23.92}}} & {\ul{25.47}} & \multicolumn{1}{c}{{\ul{23.95}}} & {\ul{25.25}} & \multicolumn{1}{c}{{\ul{23.51}}} & {\ul{25.29}} \\ 
\multicolumn{1}{c}{} & \begin{tabular}[c]{@{}l@{}}Proposed \end{tabular} & \multicolumn{1}{c}{{\textbf{\emph{28.47}}}} & {\textbf{\emph{28.32}}} & \multicolumn{1}{c}{{\textbf{\emph{28.63}}}} & {\textbf{\emph{28.42}}} & \multicolumn{1}{c}{{\textbf{\emph{28.61}}}} & {\textbf{\emph{28.26}}} & \multicolumn{1}{c}{{\textbf{\emph{28.88}}}} & {\textbf{\emph{28.27}}} & \multicolumn{1}{c}{{\textbf{\emph{28.76}}}} & {\textbf{\emph{28.40}}} \\ 
\multicolumn{1}{c}{\multirow{3}{*}{Density is 0.1 }} & \cite{luo2019adaptive} & \multicolumn{1}{c}{20.09} & 21.86 & \multicolumn{1}{c}{19.83} & 21.69 & \multicolumn{1}{c}{19.92} & 21.92 & \multicolumn{1}{c}{19.59} & 21.34 & \multicolumn{1}{c}{19.86} & 21.95 \\ 
\multicolumn{1}{c}{} & \cite{zeng2021quantum} & \multicolumn{1}{c}{{\ul{20.81}}} & {\ul{22.47}} & \multicolumn{1}{c}{{\ul{20.74}}} & {\ul{22.56}} & \multicolumn{1}{c}{{\ul{21.24}}} & {\ul{22.59}} & \multicolumn{1}{c}{{\ul{21.35}}} & {\ul{22.48}} & \multicolumn{1}{c}{{\ul{20.81}}} & {\ul{22.28}} \\ 
\multicolumn{1}{c}{} & \begin{tabular}[c]{@{}l@{}}Proposed \end{tabular} & \multicolumn{1}{c}{{\textbf{\emph{23.88}}}} & {\textbf{\emph{24.02}}} & \multicolumn{1}{c}{{\textbf{\emph{23.89}}}} & {\textbf{\emph{24.01}}} & \multicolumn{1}{c}{{\textbf{\emph{23.89}}}} & {\textbf{\emph{24.09}}} & \multicolumn{1}{c}{{\textbf{\emph{23.71}}}} & {\textbf{\emph{23.98}}} & \multicolumn{1}{c}{{\textbf{\emph{23.94}}}} & {\textbf{\emph{24.05}}} \\  \hline
\end{tabular}
\label{comp_noise}
\end{table*}

\section{Discussion}
This paper investigated quantum watermarking methods algorithmically and experimentally. We have demonstrated a new AQSM model for quantum watermarking processing and HDWM, an embedding strategy based on quantum watermarking features, followed by a quantum watermarking scheme resistant to noise and cropping attacks, which has been experimentally simulated via MATLAB. The experimental results show that for different sizes of watermark images, our quantum watermarking method not only made the embedded watermark imperceptible, but also showed good robustness against high-density noise and large-area cropping attacks, which was consistent with our expectation. In addition, experimental data show that the effect of the embedded watermark on the carrier image was negatively correlated with the robustness of the watermark in the quantum watermarking algorithm, a result that is in line with previous research results. The research in this paper was precisely the result of the trade-off between image quality and robustness, and the robustness was improved as much as possible while preserving the image quality.

Compared with previous studies, especially those based on spatial-domain quantum watermarking, we confirm that adaptive embedding of watermarks of different sizes into the same carrier image is feasible. This has not been mentioned in most studies. In addition, experimental results demonstrate the advantage of our robustness. Our study provides new algorithms and experimental evidence for quantum watermarking methods, which is useful for subsequent research. 

\section{Conclusion}\label{part6}
We have presented the AQSM for watermark processing and HDWM for watermarking. Unlike the existing quantum watermarking schemes with fixed embedding scales, the proposed quantum watermarking scheme can flexibly embed watermarks with different scales and has a better robustness. In order to improve the accuracy of extracting watermark information, a quantum refining method is proposed that enables the extraction of logical information from a sequence of odd qubits with error. Finally, the effectiveness and robustness of the proposed quantum watermarking method are evaluated. Three different embedding ratios (1;4,1:16,1:64) of the watermarking algorithm are simulated using the database USC-SIPI as the carrier images. The results show that our watermarking scheme has good invisibility and robustness. In particular, it can cope with high-density noise addition and large-area cropping attacks when the scale factor $r$ is greater than 2.

\section*{Acknowledgements}
This work was funded by the Macao Polytechnic University (Project No. RP/FCA-04/2024), and the Science and Technology Development Fund of Macau SAR (Grant number 0045/2022/A).


\small
\bibliographystyle{IEEEtran}
\bibliography{ref}

\end{document}